\documentclass[twocolumn]{aastex62}
\usepackage{amsmath}
\usepackage{graphicx}
\usepackage{color}
\usepackage{amsmath}
\usepackage{graphicx}

\newcommand{\beq}{\begin{equation}}
\newcommand{\eeq}{\end{equation}}

\newcommand{\dd}{\partial}

\begin{document}

\title{FANTASY: User-Friendly Symplectic Geodesic Integrator for Arbitrary Metrics with Automatic Differentiation}

\author{Pierre Christian}
\affiliation{Physics Department, Fairfield University, 1073 North Benson Road, Fairfield, CT 06824, USA}
\affiliation{Astronomy Department, University of Arizona, 933 North Cherry Avenue, Tucson, AZ 85721, USA}
\author{Chi-kwan Chan} 
\affiliation{Astronomy Department, University of Arizona, 933 North Cherry Avenue, Tucson, AZ 85721, USA}
\affiliation{Data Science Institute, University of Arizona, 1230 N. Cherry Avenue, Tucson, AZ 85721, USA}
\affiliation{Program in Applied Mathematics, University of Arizona, 617 N. Santa Rita, Tucson, AZ 85721, USA}

\begin{abstract}
We present FANTASY (Finally A Numerical Trajectory Algorithm both Straightforward and sYmplectic), a user-friendly, open-source symplectic geodesic integrator written in Python. FANTASY is designed to work "out-of-the-box" and does not require anything from the user aside from the metric and the initial conditions for the geodesics. FANTASY efficiently computes derivatives up to machine precision using automatic differentiation, allowing the integration of geodesics in arbitrary space(times) without the need for the user to manually input Christoffel symbols or any other metric derivatives. Further, FANTASY utilizes a Hamiltonian integration scheme that doubles the phase space, where two copies of the particle phase space are evolved together. This technique allows for an integration scheme that is both explicit and symplectic, even when the Hamiltonian is not separable. FANTASY comes prebuilt with second and fourth order schemes, and is easily extendible to higher order schemes. FANTASY also includes an automatic Jacobian calculator that allows for coordinate transformations to be done automatically. \\
\end{abstract}

\section{Introduction}
In the past, geodesic integrators have been used to ray trace photon orbits around black hole spacetimes. Typically, such calculations are done on top of general relativistic magnetohydrodynamics simulations in post-processing. While the majority of past integrators focus on the Kerr metric \citep[e.g.,][]{2009ApJ...696.1616D, 2016MNRAS.462..115D,2015ApJS..218....4C,2011MNRAS.410.1052S,2013ApJ...777...13C,2018ApJ...867...59C} or the Kerr-Newman metric \citep{2014A&A...561A.127Y}, the possibility of testing modified theories of gravity through electromagnetic observations of supermassive black holes at horizon scales \citep{2010ApJ...718..446J} via experiments such as the Event Horizon Telescope (EHT) \citep{PaperI,PaperII,PaperIII,PaperIV,PaperV,PaperVI} generates an impetus for the development of geodesic integrators that are capable of handling more general spacetimes. To this end, geodesic integrators have been developed as part of ray tracing algorithms that can be employed on arbitrary spacetimes \citep{2018A&A...613A...2B,2016PhRvD..94h4025Y} or on specific parametrized spacetimes \citep{2012ApJ...745....1P}. These integrators, however, are non-symplectic and thus possess unbounded errors in their conserved quantities. Comparatively, numerical methods that respect the symplectic nature of general relativistic geodesics have been relatively unexplored, with just one recent result showing that if a $(3+1)$ form of the spacetime can be supplied, an intrinsic integration scheme that preserves the Hamiltonian to numerical accuracy can be employed \citep{BRCS}. 

The parametrized spacetimes relied upon in many contemporary tests of modified gravity are phenomenological metrics that are parameterized deviations of the Schwarzschild or Kerr solutions in general relativity \citep{Vigelandmetric, JPmetric, Rezzollametric}. These metrics are not solution of any field equations, and might contain pathologies such as Lorentz violations or closed timelike curves \citep{Johannsen}. Further, many of these metrics possess geodesic equations that lack the fourth constant of motion, and thus are non-integrable. Due to their constructions, these metrics also have long and unwieldy mathematical forms. All of these difficulties make their geodesic structure relatively under-explored.

In addition to the phenomenological metrics, there are also a variety of metrics that are solutions to particular modified theories of gravity whose geodesics remain largely unplumbed. Some examples include neutron stars in scalar-tensor gravity \citep{1992ApJ...393..685Z, PhysRevLett.70.2220} and $f(R)$ gravity \citep{2010PhRvD..82f4033C,2009PhRvD..80f4002U}, black holes in Chern-Simons gravity \citep{2009PhRvD..79h4043Y,2009PThPh.122..561K}, as well as configurations consisting of combinations of black holes \citep{2008LRR....11....6E} and black rings \citep{2008LRR....11....6E} in higher dimensional theories \citep{2007JHEP...05..050E}.

In this work we present FANTASY (Finally A Numerical Trajectory Algorithm both Straightforward and sYmplectic), an open-source geodesic integrator that is specifically written to be user-friendly. In particular, FANTASY only takes as input the metric and the initial conditions for the geodesic. It does not require the user to input Christoffel symbols or any derivatives of the metric, as it employs automatic differentiation to efficiently compute metric derivatives to machine precision. As it directly integrates the geodesic equation, the user also does not need to manually supply any conserved quantities. FANTASY is capable of handling metrics that are not splittable to a $(3+1)$ form, and thus can be used to integrate geodesics in non-globally hyperbolic spacetimes and Riemannian spaces. The former is useful for integrating geodesics in designer metrics containing pathologies \citep{Johannsen}, while the latter is useful for integrating geodesics in statistical manifolds equipped with the Fisher information metric \citep{2007PhRvL..99j0602C,2017arXiv170807211I}. The integration scheme we employed is symplectic, thus allowing only for bounded errors in the conserved quantities. FANTASY comes prebuilt with second and fourth order integration schemes, and can be easily extended to higher order (even) schemes.

In Section \ref{sec:geoin} we describe geodesic integration in general, in Section \ref{sec:scheme} we discuss the symplectic integration scheme used by FANTASY, and in Section \ref{sec:autodif} we present a formulation of automatic differentiation that is suitable for geodesic integration in arbitrarily curved manifolds. We follow these technical discussions with an application of FANTASY to the Kerr metric in Section \ref{sec:Kerr} and the Kerr-Sen metric \ref{sec:Kerr-Sen}. Finally, in Section \ref{sec:conclude} we give some concluding remarks.

Unless specified otherwise, in this paper we will work on an $n$ dimensional manifold with a metric signature (-,+,+,+,\ldots,+). In Darboux coordinates, our $2n$-phase space coordinates will be $(q,p)$, where coordinates without indices indicate the collection $q=\{q^0, q^1, \ldots, q^n \}$.

\section{Geodesic Integration on Manifolds} \label{sec:geoin}
The Lagrangian for geodesics in curved spaces (or spacetimes) is
\beq \label{eq:Lagrangian}
L(q,u) = \frac{1}{2} g_{\alpha \beta} u^\alpha u^\beta \; ,
\eeq
where $u^\alpha = \mathrm{d} q^\alpha/ \mathrm{d} \lambda$ is the rate of change with respect to the affine parameter $\lambda$. Performing the Legendre transformation we obtain the Hamiltonian
\beq \label{eq:Hamiltonian}
H(q,p) = \frac{1}{2} g^{\alpha \beta} p_\alpha p_\beta \; ,
\eeq
as a function of the conjugate momenta $p_\alpha$. In curved manifolds, it is important to note that the Hamiltonian is defined on the cotangent bundle instead of the tangent bundle. This is because components of vectors in the tangent space are related to components of covectors on the cotangent bundle by the metric, and components of the metric in general change with the coordinates and thus possess non-zero derivatives. 

Once written in the Hamiltonian form, we can solve for the geodesics using standard Hamiltonian dynamics defined on the cotangent bundle, now seen as a $2n$ dimensional symplectic manifold with symplectic 2-form $\Omega = \mathrm{d}p \wedge \mathrm{d}q$ in local Darboux coordinates. In these coordinates, the Hamilton equations read
\begin{align}
\frac{\mathrm{d}q^\alpha}{d\lambda} &= \frac{\partial H}{\partial p_\alpha} \; ,
\\ \frac{\mathrm{d}p_\alpha}{d\lambda} &= -\frac{\partial H}{\partial q^\alpha} \; .
\end{align}
For $H$ given by equation \ref{eq:Hamiltonian}, the Hamilton equations become,
\begin{align}
\frac{\mathrm{d}q^\alpha}{d\lambda} &= g^{\alpha \beta} p_\beta\; ,
\\ \frac{\mathrm{d}p_\alpha}{d\lambda} &= -\frac{1}{2} p_\gamma p_\beta \frac{\dd g^{\gamma \beta} }{\dd q^\alpha} \; . \label{eq:Ham2}
\end{align}
FANTASY solves these coupled equations through an explicit symplectic scheme aided by automatic differentiation to compute the derivatives of the metric appearing in equation (\ref{eq:Ham2}). 

\section{Explicit and Symplectic Integration Scheme} \label{sec:scheme}
FANTASY uses an explicit, symplectic integration scheme based on embedding the phase space, with coordinates $(q,p)$ into a "doubled" phase space consisting of two copies of the phase space,
$$
(q,p) \rightarrow (q,p,x,y) \; ,
$$
where $(x,y)$ are coordinates for the second copy of the phase space. In local coordinates, the doubled phase space is endowed with the symplectic 2-form $\mathrm{d}q \wedge \mathrm{d}p + \mathrm{d}x \wedge \mathrm{d}y$. This method was first proposed by \cite{Pihajoki} for non-symplectic integrators and extended by \cite{Tao} for symplectic integrators. A symplectic integration scheme ensures that the conserved quantities of the orbit possess bounded errors, however note that it does not ensure that the errors are at machine precision.

In order to integrate the flow of a Hamiltonian $H$, we instead solve the flow of $\bar{H}$, a Hamiltonian of the doubled phase space given by,
\beq
\bar{H} (q,p,x,y) = H_A + H_B + H_C \; ,
\eeq
where $H_A = H(q,y)$ and $H_B = H(x,p)$ are the original Hamiltonian with the position and momenta coordinates mixed between the first and second copies of the phase space, $H_C = (\omega/2) ( ||q-x||^2_2 + ||p-y||^2_2) $ a constraint Hamiltonian that couples the two copies of the phase space, and $\omega$ a scalar parameterizing the strength of the coupling.

The flow of $\bar{H}$ on the doubled phase space is then constructed by Strang splitting the flow map of $\bar{H}$. For a timestep $\delta$, 
\beq \label{map}
\phi^\delta_{\bar{H}} = \phi_{H_A}^{\delta/2} \circ \phi_{H_B}^{\delta/2} \circ \phi_{H_C}^\delta \circ \phi_{H_B}^{\delta/2} \circ \phi_{H_A}^{\delta/2} \; , 
\eeq
where $\phi_Y^T$ is the flow map of a hamiltonian $Y$ over timestep $T$. This integration scheme is second order in accuracy, and as shown in \cite{Tao}, is explicit and symplectic, even for non-separable Hamiltonians.  
Higher order integrators can then be constructed out of the second order flow map, $\phi^\delta_{\bar{H}}$ by the Yoshida triple-jump \citep{yoshida},
\beq
\phi^{\delta}_{2n+2} = \phi^{z_1 \delta}_{2n} \circ  \phi^{z_0 \delta}_{2n}  \circ  \phi^{z_1 \delta}_{2n} \; ,
\eeq
where $\phi^{\delta}_{2n+2}$ and $\phi^{\delta}_{2n}$ are the order $2n+2$ and order $2n$ flow maps with timestep $\delta$, respectively,  
\beq
z_0 \equiv -\frac{2^{1/(2n+1)}}{2 - 2^{1/(2n+1)}} \;, 
\eeq 
and
\beq
z_1 \equiv \frac{1}{2-2^{1/(2n+1)}} \;.
\eeq
FANTASY comes prebuilt with the second and fourth order integration schemes.

\section{Automatic Differentiation} \label{sec:autodif}

FANTASY performs differentiations to machine precision automatically through the use of dual numbers. Dual numbers have been used to perform automatic differentiations in the machine learning literature, where they are treated as ad hoc extensions of the real numbers. Here we provide a dual number formalism in the language of exterior algebra. This formalism allows for an easy generalization to multivariable calculus, as well as extensions to arbitrarily curved manifolds. 

\subsection{Dual numbers in one dimension}

The \emph{dual number} in one dimension is an extension of $\mathbb{R}$ to $\mathbb{R}^2$ much like the complex number. For complex numbers, the second copy of $\mathbb{R}$ is multiplied by the complex element $i$, with the property $i^2 = -1$. For dual numbers, the second copy of $\mathbb{R}$ is multiplied by the dual element $\mathrm{d}x$, with the property $\mathrm{d}x^2 = 0$. A dual number $D$ can therefore be written as 
\beq
D = (a + b \mathrm{d}x) \; ,
\eeq
where both $a$ and $b$ are real numbers. For two dual numbers $D= (a + b \mathrm{d}x)$ and $G=(c + d \mathrm{d}x)$ dual numbers, their multiplication is given by,
\beq \label{eq:derivatives}
DG = ac + (ad + bc) \mathrm{d} x \; .
\eeq
As our notation suggests, the dual numbers are members of the exterior algebra, $\bigwedge V \equiv T(V)/I$, i.e., the algebra of the wedge products, constructed by the quotient of the tensor algebra,
\beq
T(V) = \mathbb{R} \oplus V \oplus (V \otimes V) \oplus  (V \otimes V \otimes V) \ldots  \; ,
\eeq
for a one-dimensional vector space $V$, with the two-sided ideal $I$ consisting of the set of elements of the form
\beq
D ( \mathrm{d}x \otimes \mathrm{d}x) G \; ,
\eeq
where $D$ and $G$ members of $T(V)$ and $\otimes$ the usual tensor product. 

Next, we define the \emph{dual number form} of a real function of one variable, $f(x)$, as the following,
\beq
F \equiv f + f' \mathrm{d}x \; . 
\eeq
This is a map between real functions of one variable to the dual numbers. Notice that the multiplicative algebra of dual numbers given by equation (\ref{eq:derivatives}) automatically returns the product rule,
\beq
FH = fh + (fh' + hf') \mathrm{d}x \; ,
\eeq
for $f$, $h$ arbitrary real functions of one variable, and where $H= h + h' \mathrm{d}x$ is the dual number form of $h$. The case for additions is also readily apparent,
\beq
F + H = f+h + (f'+h') \mathrm{d}x \; .
\eeq
As such, by writing all of our real functions in their dual number forms, and performing all our algebra in dual number space, we get for free the derivatives of all the multiplications and additions of our functions. This means that we automatically calculate the derivatives of all analytic functions. 

This automatic differentiation scheme can be further augmented by defining the action of common functions on dual numbers. For example, 
\begin{align}
\sin &: \bigwedge V \rightarrow \bigwedge V \;, \nonumber
\\ &\;\;\; a+b \mathrm{d}x \rightarrow \sin(a)+ b \cos(a) \mathrm{d}x \; . \nonumber
\end{align}
Applying this to the dual number form of $f(x)$ gives 
\beq
\sin(F) = \sin(f) + f' \cos(f) \mathrm{d}x \; ,
\eeq
which again gives the derivative of $f$ in the second entry. While these additional definitions are in principle not necessary as long as all the functions of interests are analytic, in practice it speeds up computations due to common functions being highly optimized in languages such as Python.

\subsection{Dual numbers in general phase-spaces}

We can generalize the dual number for an $n$-dimensional vector space $V$ by demanding again that they are members of the exterior algebra, $T(V)/I$, where now $V$ is an $n$-dimensional vector space and $I$ is the two sided ideal consisting of the set of elements of the form 
\beq
\sum_i^n D_i ( \mathrm{d}x^i \otimes \mathrm{d}x^i) G_i \; ,
\eeq
where $D_i$ and $G_i$ are members of $T(V)$. For the coordinates $\{x^1,x^2,\ldots,x^n\}$, a typical dual number has the form,
\begin{align}
D &= a_{1,1} \mathrm{d}x^1 + a_{2,1} \mathrm{d}x^2 + a_{31} \mathrm{d}x^3 + \ldots + a_{n,1} \mathrm{d}x^n \nonumber
\\ &+ a_{1,2} \mathrm{d}x^1 \wedge \mathrm{d}x^2 + a_{1,3} \mathrm{d}x^1 \wedge \mathrm{d}x^3 + \ldots + a_{1,n} \mathrm{d}x^1 \wedge \mathrm{d}x^n \nonumber
\\ &+ \ldots \nonumber
\\ &+ a_{n,1} \mathrm{d}x^n \wedge \mathrm{d}x^1 + \ldots +  a_{n,(n-1)} \mathrm{d}x^n \wedge \mathrm{d}x^{(n-1)} \; , \nonumber
\end{align} 
where $a_{1,2}$ are real numbers and $\wedge$ is the usual exterior product. 

We define the \emph{dual number form} of a real function of $n$-variables, $f(x^1,x^2,\dots,x^n)$, as 
\beq
F \equiv f + \frac{\dd f}{\dd x^1} \mathrm{d}x^1 + \frac{\dd f}{\dd x^2} \mathrm{d}x^2 + \ldots + \frac{\dd f}{\dd x^n} \mathrm{d}x^n \; .
\eeq
This is a map between a function of $n$-variables to instances of dual numbers with no $\mathrm{d}x^i \wedge \mathrm{d}x^j$ terms. As can be easily verified, multiplying the dual number forms of two $n$-dimensional functions $f$ and $h$ automatically gives us the product rule, 
\begin{align}
FH =& fh \nonumber
\\&+ \left(f \frac{\dd h}{\dd x^1} + h \frac{\dd f}{\dd x^1} \right) \mathrm{d}x^1 \nonumber
\\&+ \ldots \nonumber
\\&+ \left(f \frac{\dd h}{\dd x^n} + h \frac{\dd f}{\dd x^n} \right) \mathrm{d}x^n \nonumber
\\ &+ \mathrm{cross \; terms} \; ,
\end{align}
where the cross terms are of the form
\beq
\left(\frac{\dd f}{\dd x^i} \frac{\dd h}{\dd x^j} - \frac{\dd f}{\dd x^j} \frac{\dd h}{\dd x^i} \right) \mathrm{d}x^i \wedge \mathrm{d}x^j \; .
\eeq
If the vector space $V$ is a $2n$-dimensional symplectic vector space in Darboux coordinates, $\{x^1,\ldots x^n,p^1,\ldots p^n\}$, then the cross terms in the multiplication of $FG$ automatically computes the Poisson bracket,
\beq
\{f,g\} = \sum_{i=1}^n \frac{\dd f}{\dd x^i} \frac{\dd h}{\dd p^i} - \frac{\dd f}{\dd p^i} \frac{\dd h}{\dd x^i}  \; .
\eeq

The next generalization is to arbitrarily curved manifolds. Due to our dual numbers being defined using the exterior algebra formalism, this can be done in a straightforward manner. For an $n$-dimensional manifold $M$ representing our space or spacetime, its cotangent space $Q \equiv T^*M$ is a $2n$-dimensional symplectic manifold that represents our phase-space. For a point $p$ on the phase-space manifold $Q$, we define the dual number as a member of the exterior algebra on the cotangent space of $Q$ at $p$, i.e., on $T^*_p T^*M$. As $T_p^*Q$ is a vector space, our previous formalism can be directly applied on it, provided we use $T_p^*Q$ in place of $V$. These dual numbers will be members of $T(T_p^*Q)/I$, where now,
\beq
T(T_p^*Q) = \mathbb{R} \oplus T_p^*Q \oplus (T_p^*Q \otimes T_p^*Q) \oplus  \ldots  \; ,
\eeq
and $I$ is the same two-sided ideal as before, but now defined on $T(T_p^*Q)$.

\subsection{Automatic Jacobian}

Besides taking derivatives of the metric, another use-case for automatic differentiation is to perform coordinate transformations by way of computing the Jacobian. Given a coordinate transformation $q'_{i'}(q_0,q_1,\ldots)$ from the unprimed to the primed coordinates, the Jacobian,
\beq
J_{i'j} = \frac{\partial{q'_{i'}}}{\partial q_j} \; ,
\eeq
can be used to transform any tensorial quantities (such as the momenta or the metric tensor) to the primed coordinates.

Typically these derivatives have to be computed manually. FANTASY, however, has the capability for the automatic differentiation of the coordinate transform functions and thus automatic computation of the Jacobian. Combined with the coordinate transformation functions, the automatically computed Jacobian allows for tensorial quantities to be transformed in a straightforward and user-friendly manner.

\section{Geodesics of the Kerr Metric} \label{sec:Kerr}

\begin{figure*}
\centering
\includegraphics[width=2.2in]{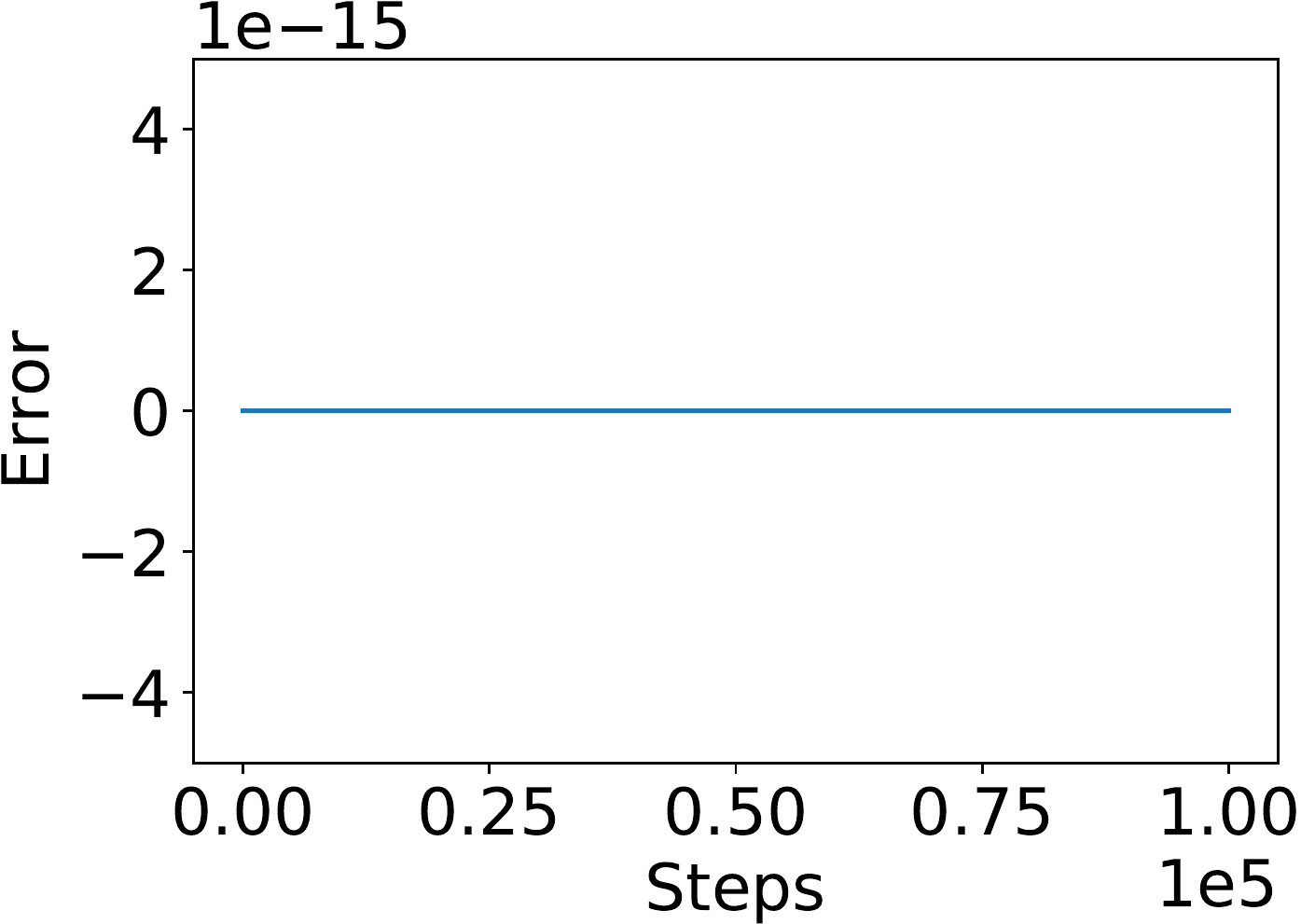} 
\includegraphics[width=2.2in]{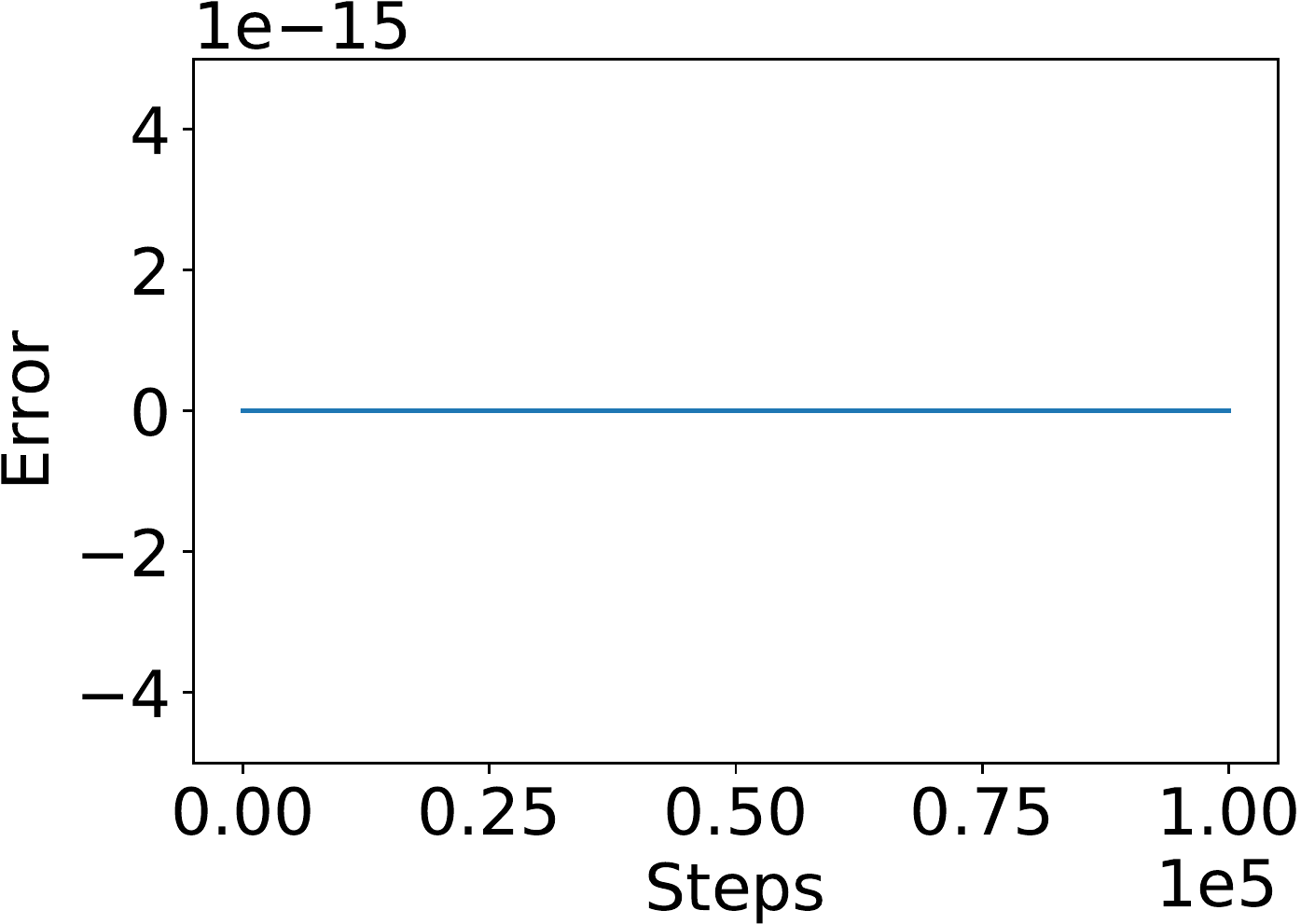}
\includegraphics[width=2.2in]{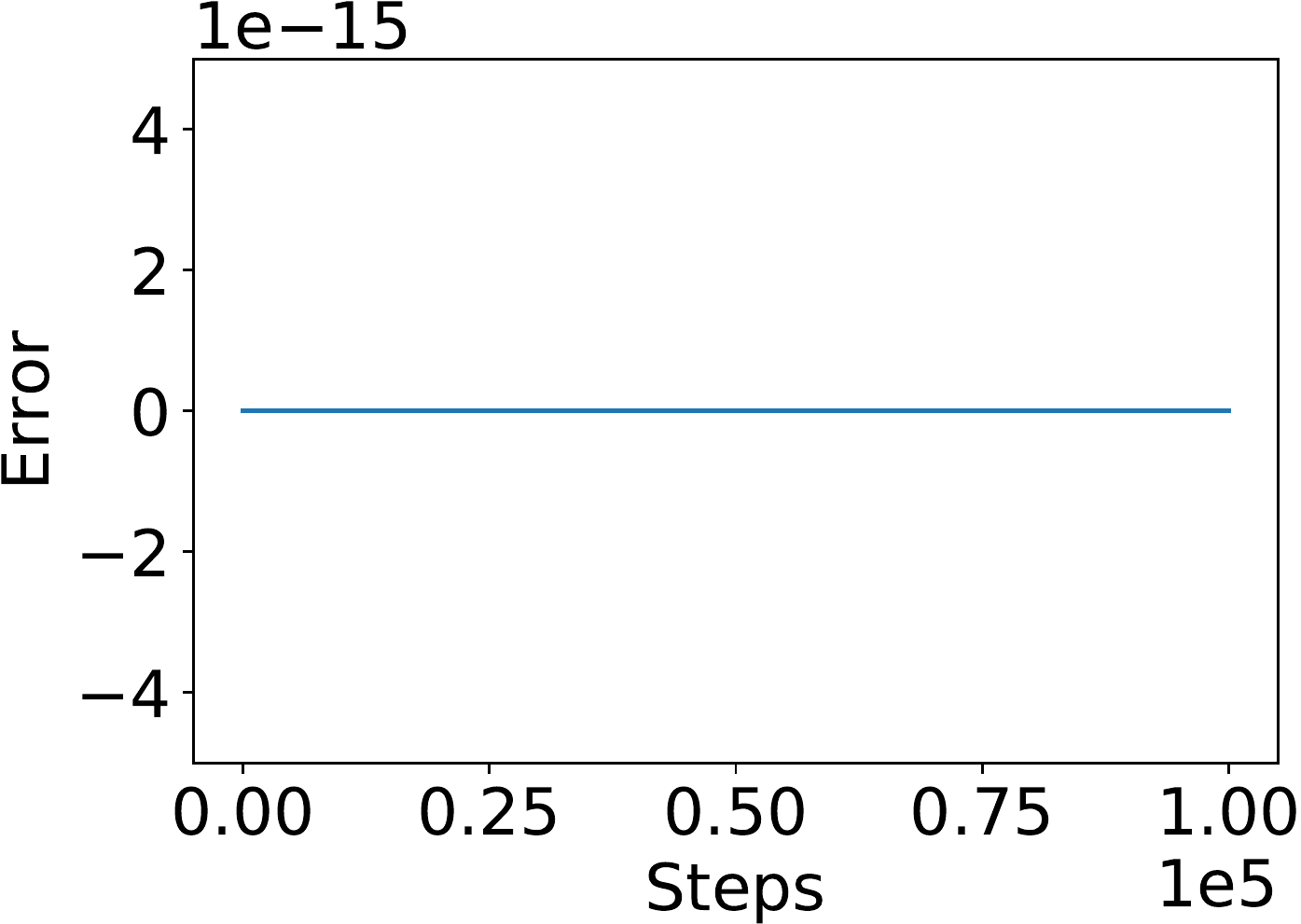} \\
\includegraphics[width=2.2in]{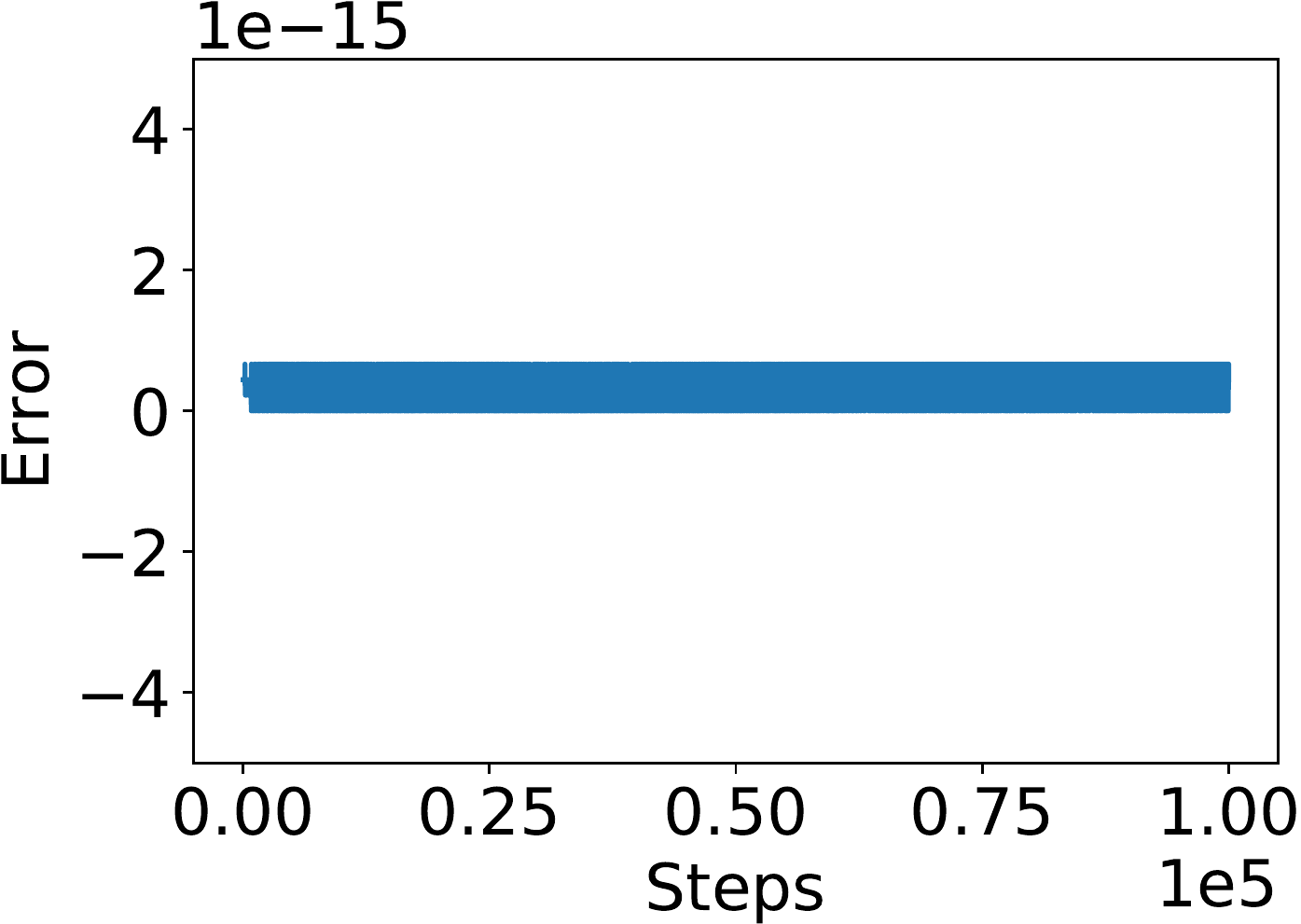} 
\includegraphics[width=2.2in]{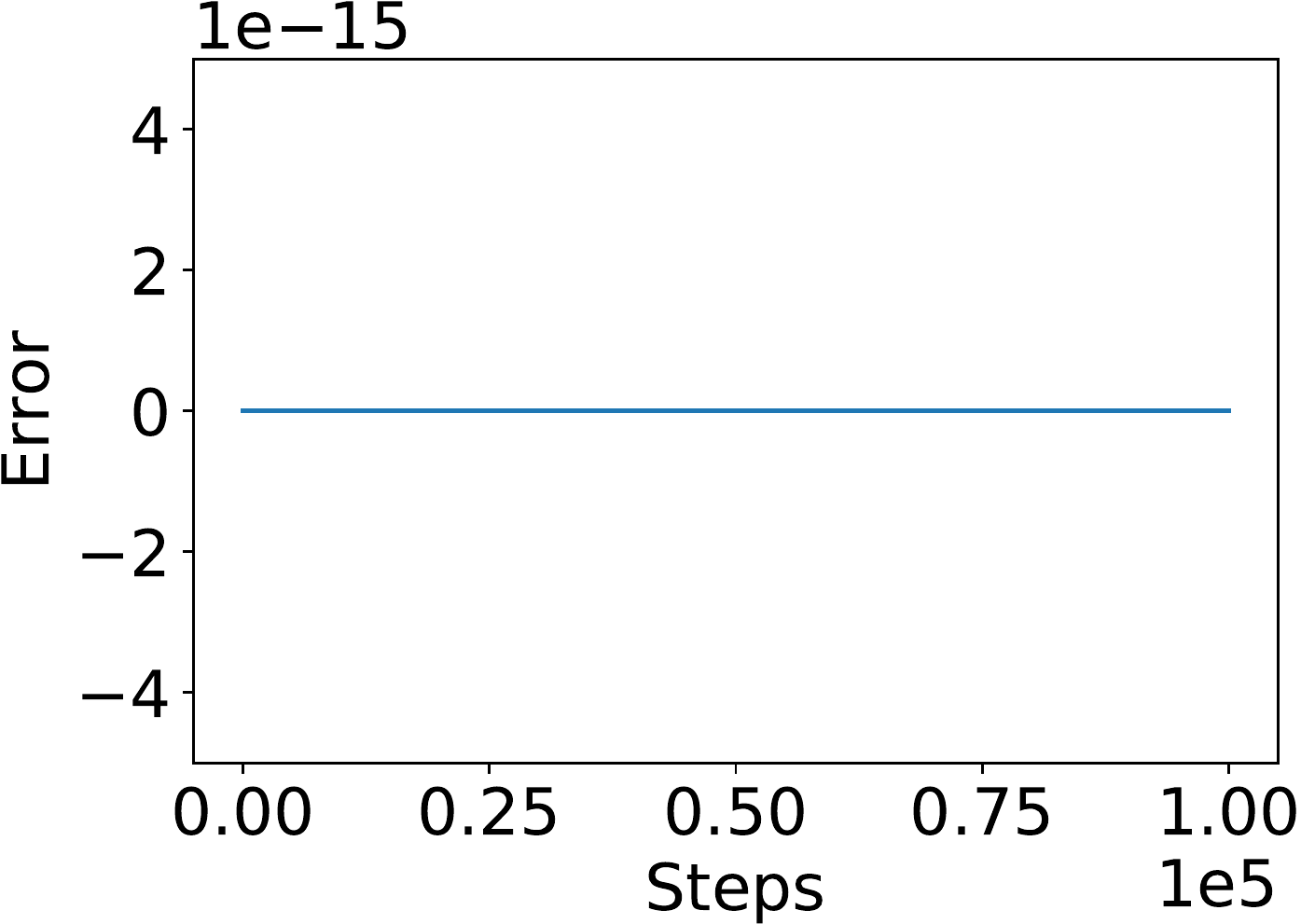}
\includegraphics[width=2.2in]{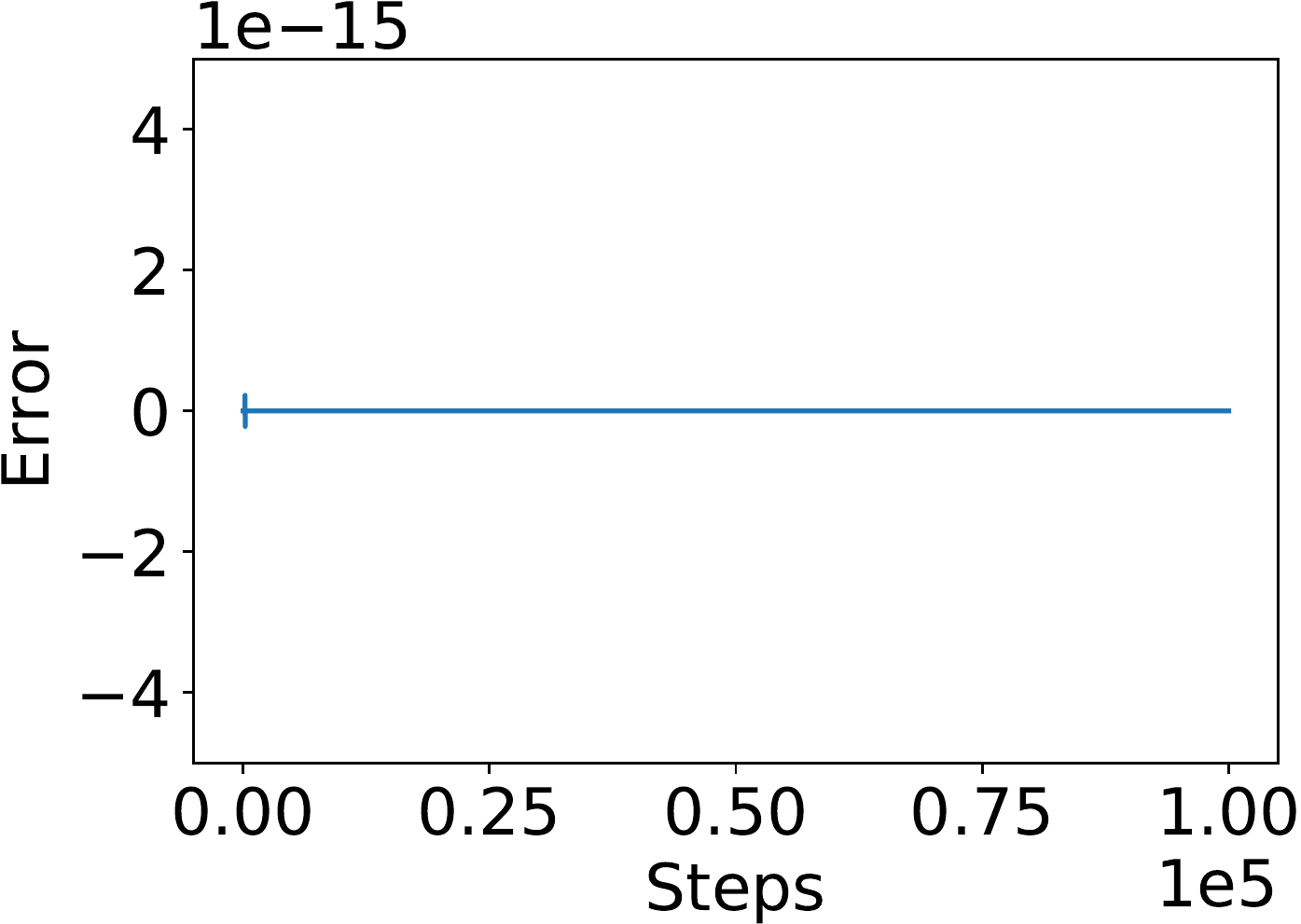}
\caption{The errors in $u \cdot u$ (left), energy (middle), and angular momentum (right) for an innermost stable circular orbit around a Schwarzschild black hole (top) and a Kerr black hole with $a=0.5$ (bottom), integrated up to $10^5$ with a timestep of with $\delta = 1$ and $\omega =1$ (top).}
\label{fig:circular}
\end{figure*}

We tested FANTASY by integrating the geodesics of the Kerr metric. The Kerr metric components are given in Boyer-Lindquist coordinates $\{ t, r, \theta, \phi \}$ by
\begin{align}
g_{tt} &= -\left( 1 - \frac{2 M r}{\rho^2} \right) \; , \\
g_{rr} &= \frac{\rho^2}{\Delta} \; , \\
g_{\theta \theta} &= \rho^2 \; , \\
g_{\phi \phi} &= \left( \rho^2 + a^2 \sin^2\theta + \frac{2 M r a^2 \sin^2\theta}{\rho^2} \right) \sin^2 \theta  \; , \\
g_{\phi t} &= - \frac{M r a}{\rho^2} \sin^2 \theta \; ,
\end{align}
with $M$ the mass of the black hole, $a$ the spin parameter, 
\begin{align}
\rho^2 &= r^2 + a^2 \cos^2 \theta \; , \\
\Delta &= r^2- 2Mr +a^2 \; .
\end{align}
For $\eta =\{1,0,0,0\}$ the timelike and $\zeta = \{0,0,0,1\}$ the spacelike Killing vectors in Boyer-Lindquist coordinates and $u$ the orbit's 4-velocity, the energy of the orbit, 
\beq
-p_t = - \eta \cdot u \; , 
\eeq
and the angular momentum of the orbit,
\beq
p_\phi = \zeta \cdot u \; ,
\eeq
should be conserved by the numerical scheme. In addition, the Kerr metric admits two Killing tensors that produce two more constants of the motions that should be conserved by a symplectic integrator. From the Killing tensor $g_{\alpha \beta}$, we obtain the condition that the normalization of the 4-velocity, $u \cdot u$, must be conserved along the geodesic. This statement means that a geodesic that starts timelike (null), remains timelike (null), and is a distinct statement from the fact that $u \cdot u$ can always be parameterized to equal $-1$ for timelike and $0$ for null geodesics. From the Killing tensor,
\beq
K^{\alpha \beta} = 2 \rho^2 l^{(\alpha} n^{\beta)} + r^2 g^{\alpha \beta} \; ,
\eeq
where $(\ldots)$ in the indices indicate symmetrization, we obtain the Carter constant, 
\beq
C \equiv K^{\alpha \beta} u_\alpha u_\beta \; ,
\eeq
as the fourth constant of motion that must be conserved by a symplectic scheme. For orbits that are confined to the equatorial plane, the Carter constant is zero, and furthermore reduces to a function of $p_t$ and $p_\phi$. As such, the conservation of the Carter constant for orbits in the equatorial plane simply follows from the conservations of $p_t$ and $p_\phi$. 
 
\subsection{Circular orbits}

\begin{figure*}[t!]
\centering
\includegraphics[width=3in]{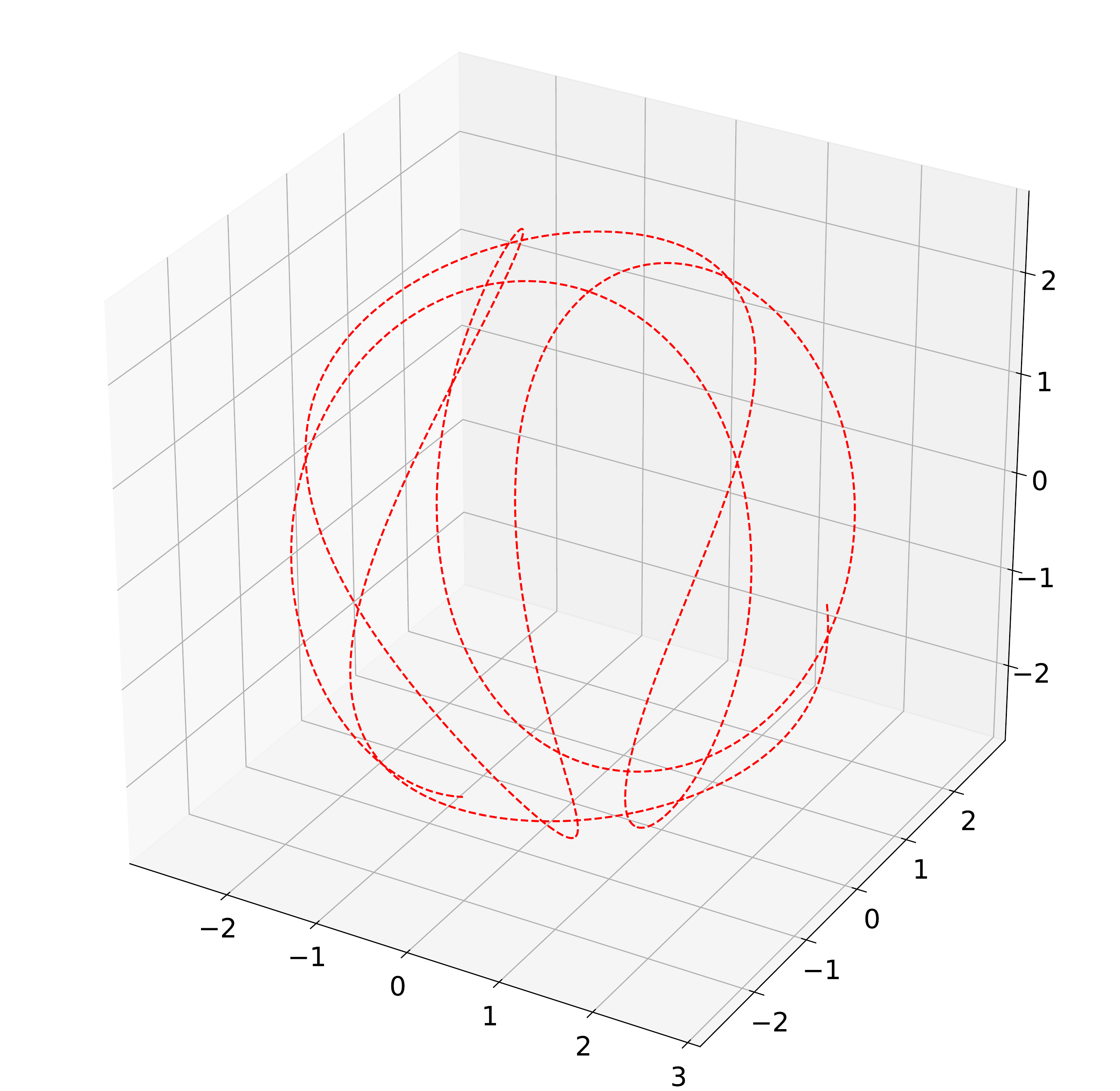}
\includegraphics[width=3in]{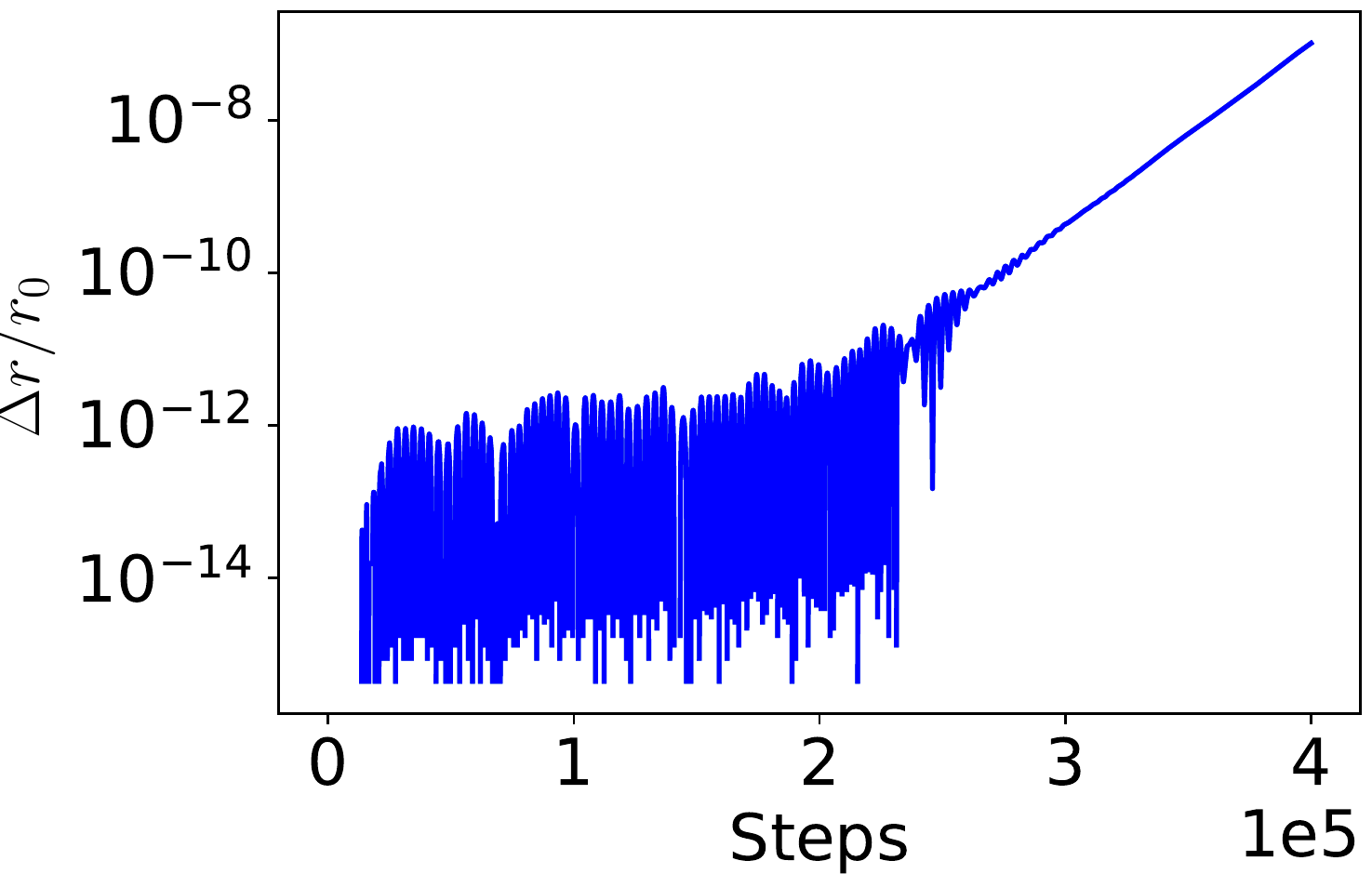} 
\caption{An unstable null geodesic around a Kerr black hole with $M=a=1$. The particles start at $q=(0,1+\sqrt{3},\pi/2,0)$ and $p=(-1,0,\sqrt{12+8 \sqrt{3}},-1) $ (left). The exponential growth in radius (right) is as expected for unstable solutions \citep{2018ApJ...867...59C}.}  
\label{fig:null_sphere}
\end{figure*}

The Kerr metric admits circular timelike geodesics in the black hole equatorial plane. If our integration scheme is symplectic, these orbits will remain circular even if integrated over many orbits. In order to obtain the initial conditions for circular orbits, first note that the Euler-Lagrange equation with equation (\ref{eq:Lagrangian}) as the Lagrangian gives
\beq
\frac{\mathrm{d}}{ \mathrm{d} \lambda} (g_{rr} \dot{r}) = \frac{1}{2} \frac{\partial g_{\alpha \beta}}{\partial r} \frac{ \mathrm{d} \dot{q}^\alpha }{\mathrm{d} \lambda} \frac{ \mathrm{d} \dot{q}^\beta }{\mathrm{d} \lambda} \; .
\eeq
Because $(\mathrm{d}r/\mathrm{d} \lambda) = (\mathrm{d}^2r/\mathrm{d}^2 \lambda) = 0 $ for circular orbits, this equation reduces to 
\beq
\frac{\partial g_{tt}}{\partial r} \left( \frac{\mathrm{d} t}{\mathrm{d} \lambda}\right)^2 + 2 \frac{\partial g_{t\phi}}{\partial r}  \frac{\mathrm{d} t}{\mathrm{d} \lambda} \frac{\mathrm{d} \phi}{\mathrm{d} \lambda} + \frac{\partial g_{\phi \phi}}{\partial r} \left( \frac{\mathrm{d} \phi}{\mathrm{d} \lambda}\right)^2 = 0 \; .
\eeq 
Solving this equation simultaneously with the timelike condition, 
\beq
\frac{ \mathrm{d}q }{ \mathrm{d} \lambda} \cdot \frac{ \mathrm{d}q }{ \mathrm{d} \lambda} = -1 \; ,
\eeq
produces a family of initial conditions for circular orbits. As a first test of FANTASY, we confirmed that the circular orbits in FANTASY remains circular over many orbital periods, and all relevant constants of motion are conserved to machine accuracy. We tested this in two cases: the innermost stable circular orbit around a Schwarzschild black hole and a circular orbit around a Kerr black hole with $a=0.5$. The errors in the constants of motion for these two tests are given in Figure \ref{fig:circular}.

\subsection{Unstable spherical null orbits}

The Kerr metric exhibits unstable spherical null orbits which provides a good testing ground for numerical geodesic integrators. Orbits in this family are given by the conditions \citep{2003GReGr..35.1909T}
\begin{align}
\frac{- p_\phi}{p_t} &= - \frac{r^3 - 3 M r^2 + a^2 r + a^2 M}{a(r - M) } \; , 
\\ \frac{C}{p_t^2} &= - \frac{r^3(r^3 - 6 M r^2 + 9 M^2 r - 4 a^2 M)}{a^2(r-M)^2}  \; . 
\end{align}
From these two equations, the initial covariant momenta $p_\alpha$ can be solved given the initial radius, $r_0$, as well as $M$ and $a$ for the black hole. To test the robustness of our integration scheme, we define
\beq
\Delta r = r-r_0 \; ,
\eeq
and plot $\Delta r/r_0$ for the case of $M=1$, $a=1$, and initial phase space coordinates $q=(0,1+\sqrt{3},\pi/2,0)$ and $p=(-1,0,\sqrt{12+8 \sqrt{3}},-1)$ in Figure \ref{fig:null_sphere}. We reproduced the exponential growth of $\Delta r/r_0$, as expected for unstable solutions \citep{2018ApJ...867...59C}.

\subsection{Generic orbits}
\begin{figure*}
\centering
\includegraphics[width=3in]{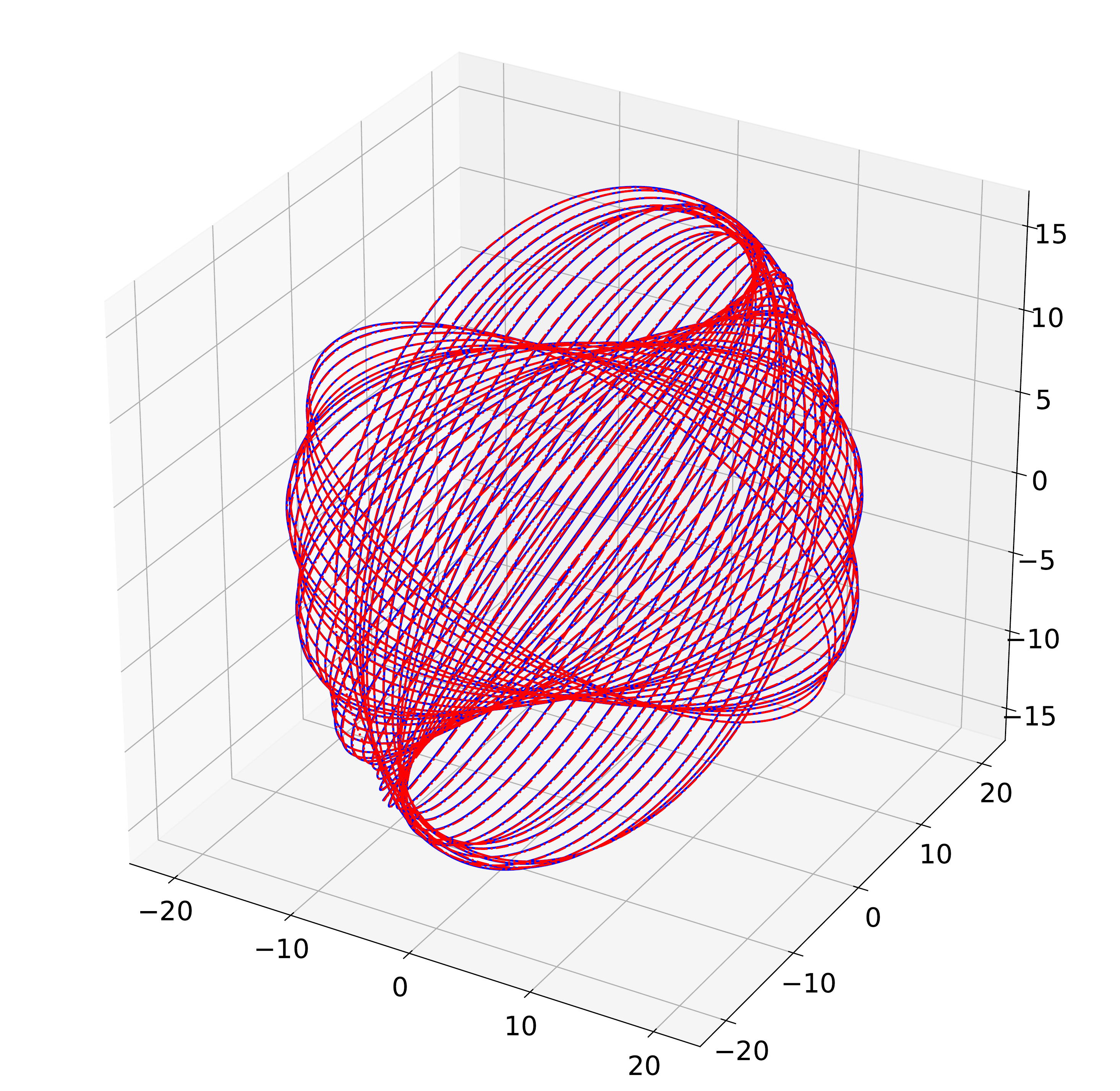} \\
\includegraphics[width=3in]{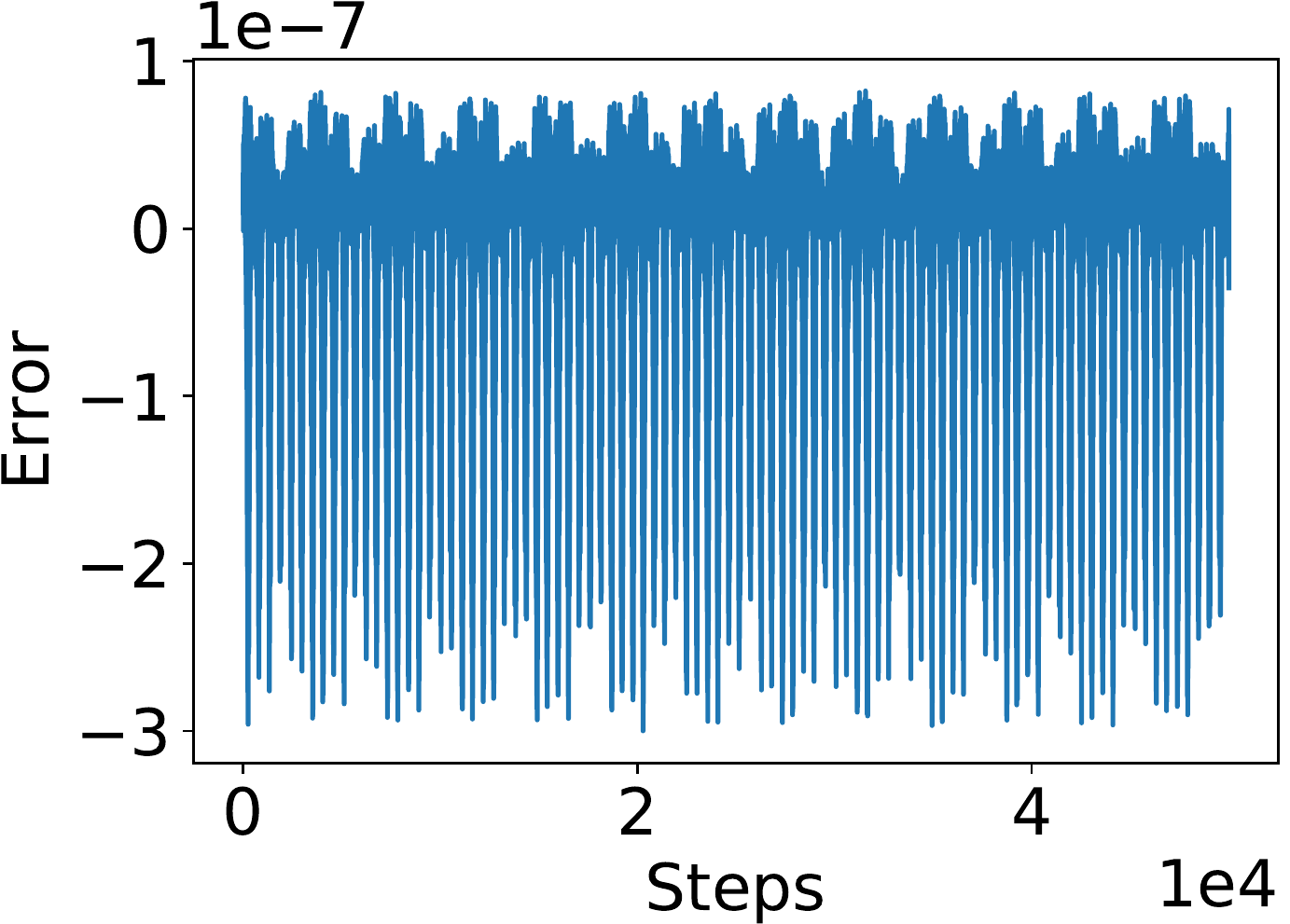} 
\includegraphics[width=3in]{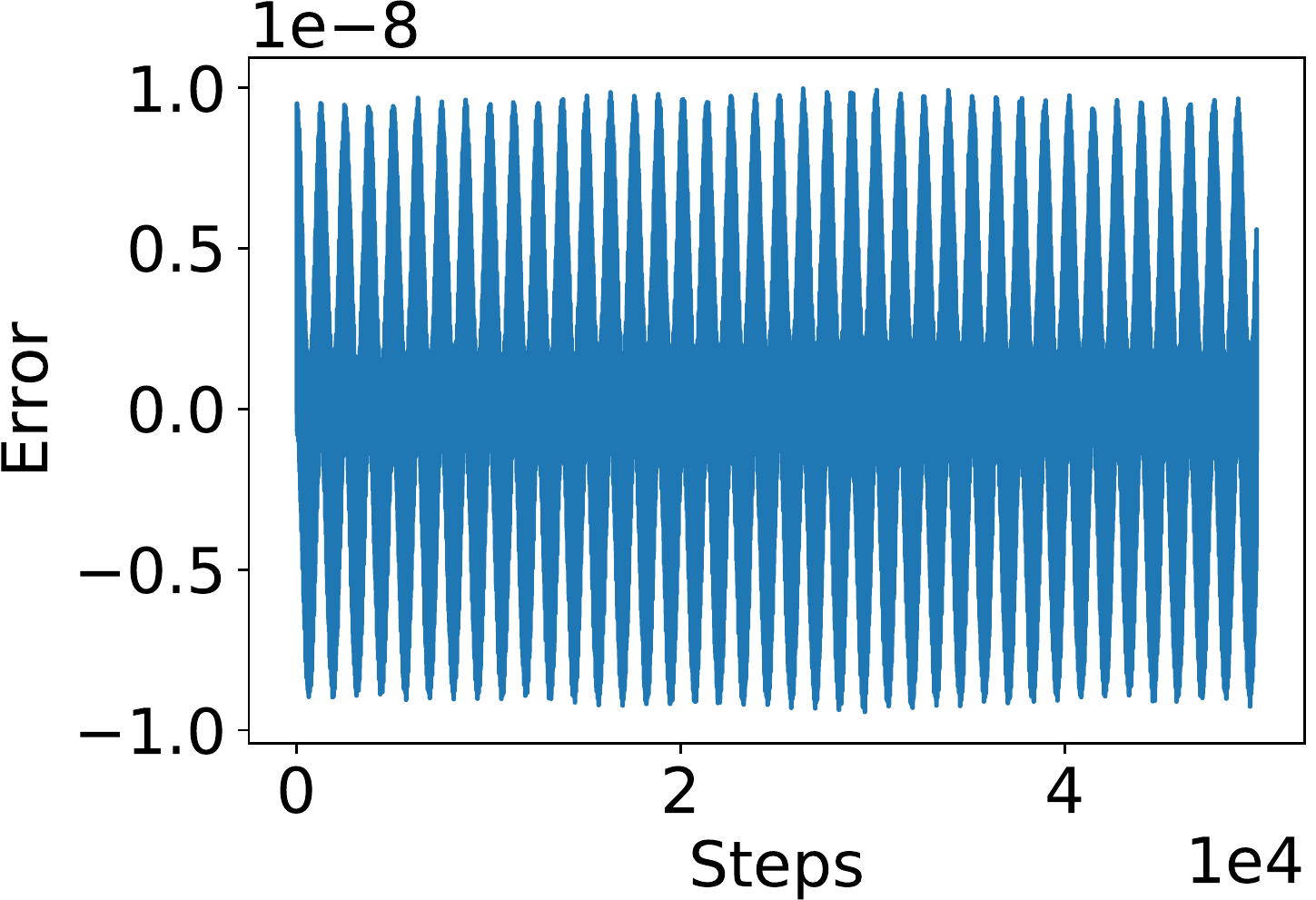}\\
\includegraphics[width=3in]{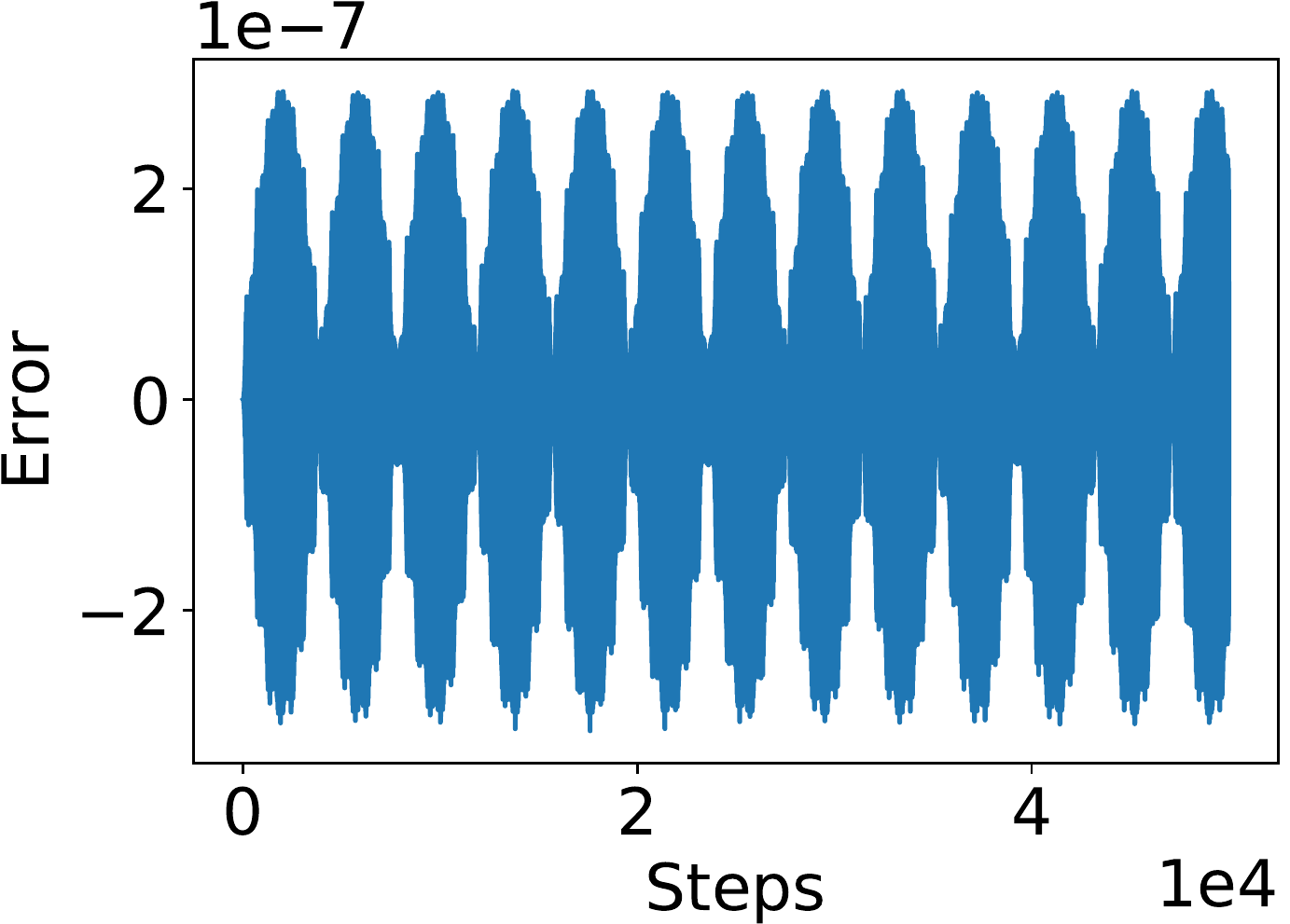}
\includegraphics[width=3in]{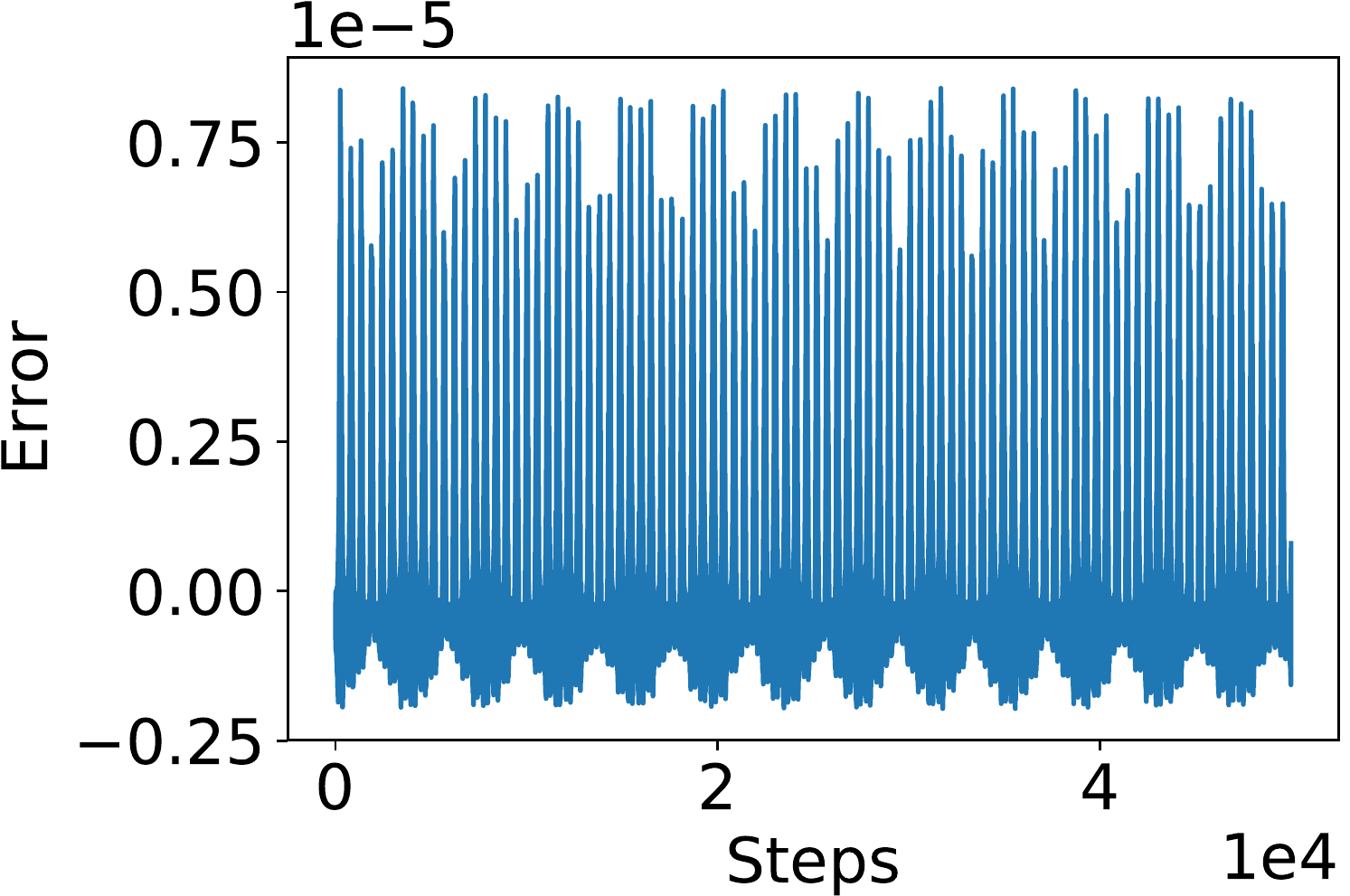}
\caption{An off-axis, non-circular orbit around a Kerr black hole with $a=0.5$, integrated up to $5 \times 10^4$ with a timestep of with $\delta = 0.5$ and $\omega =1$ (top) plotted in the KS coordinates; the orbit was computed in the Boyer-Lindquist coordinates (solid, blue) as well as the Cartesian Kerr-Schild coordinates (dot-dashed, red), and the integration results were confirmed to match. Also plotted are the errors in $u \cdot u$ (middle left), energy (middle right), angular momentum (bottom left), and Carter's constant (bottom right). The particles start at $q = (0,20,\pi/2,0)$ (black triangle) with momentum $p = (-0.9764550153430405,0,3.8,3)$.}
\label{fig:Kerr_offaxis}
\end{figure*}

\begin{figure}
\centering
\includegraphics[width=3in]{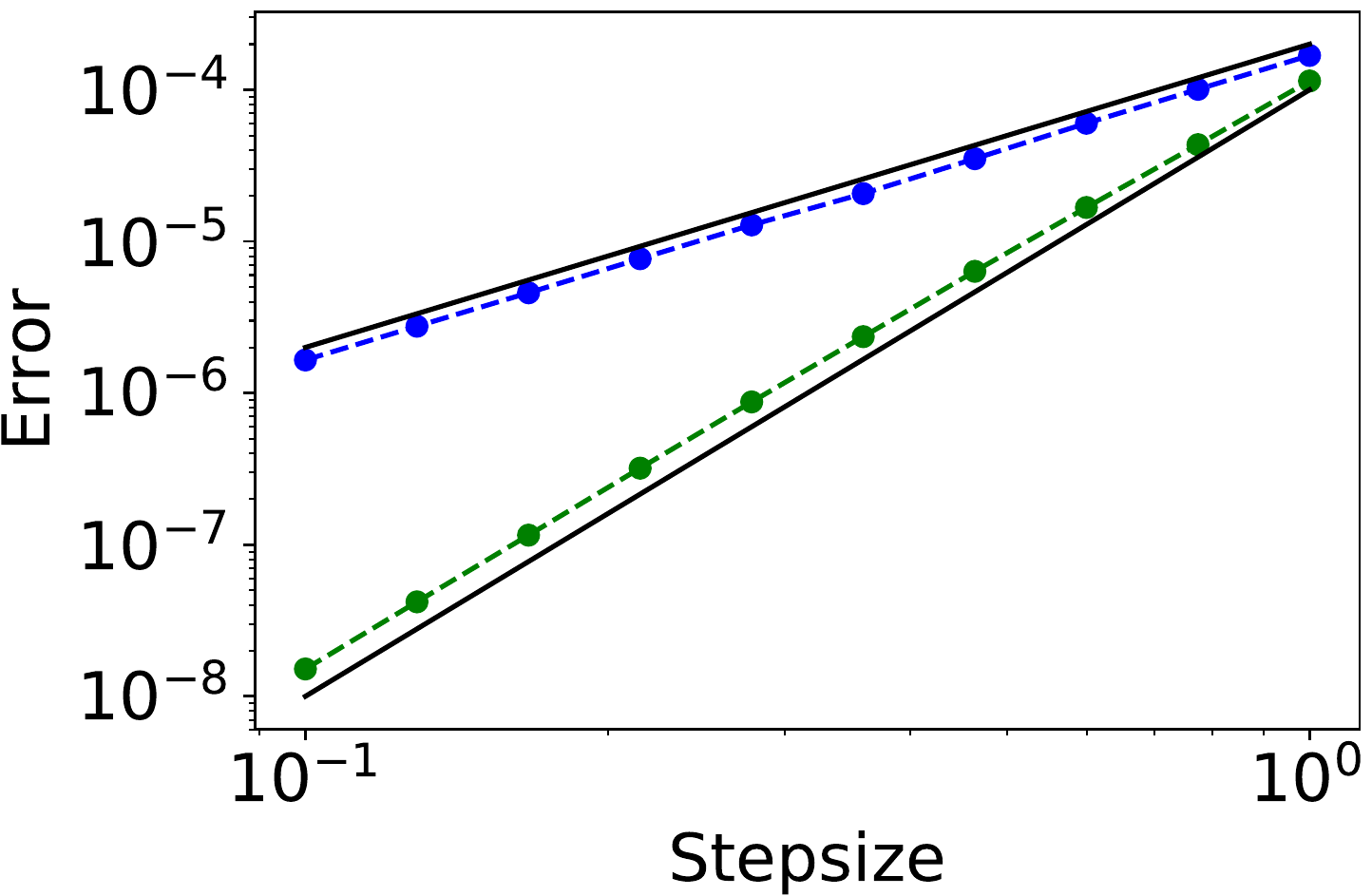} 
\caption{Error in the Carter's constant as a function of stepsize, $\Delta t$, for the second order (blue) and fourth order (green) solvers for the particles initialized around a Kerr black hole with $a=0.5$, with initial positions $q = (0,20,\pi/2,0)$ and initial momentum $p = (-0.9764550153430405,0,3.8,3)$, i.e., the same setup as in Figure \ref{fig:Kerr_offaxis}. The upper solid line is the $\propto \Delta t^2$ trendline while lower solid line is the $\propto \Delta t^4$ trendline.}
\label{fig:convergence}
\end{figure}

\begin{figure*}
\centering
\includegraphics[width=3in]{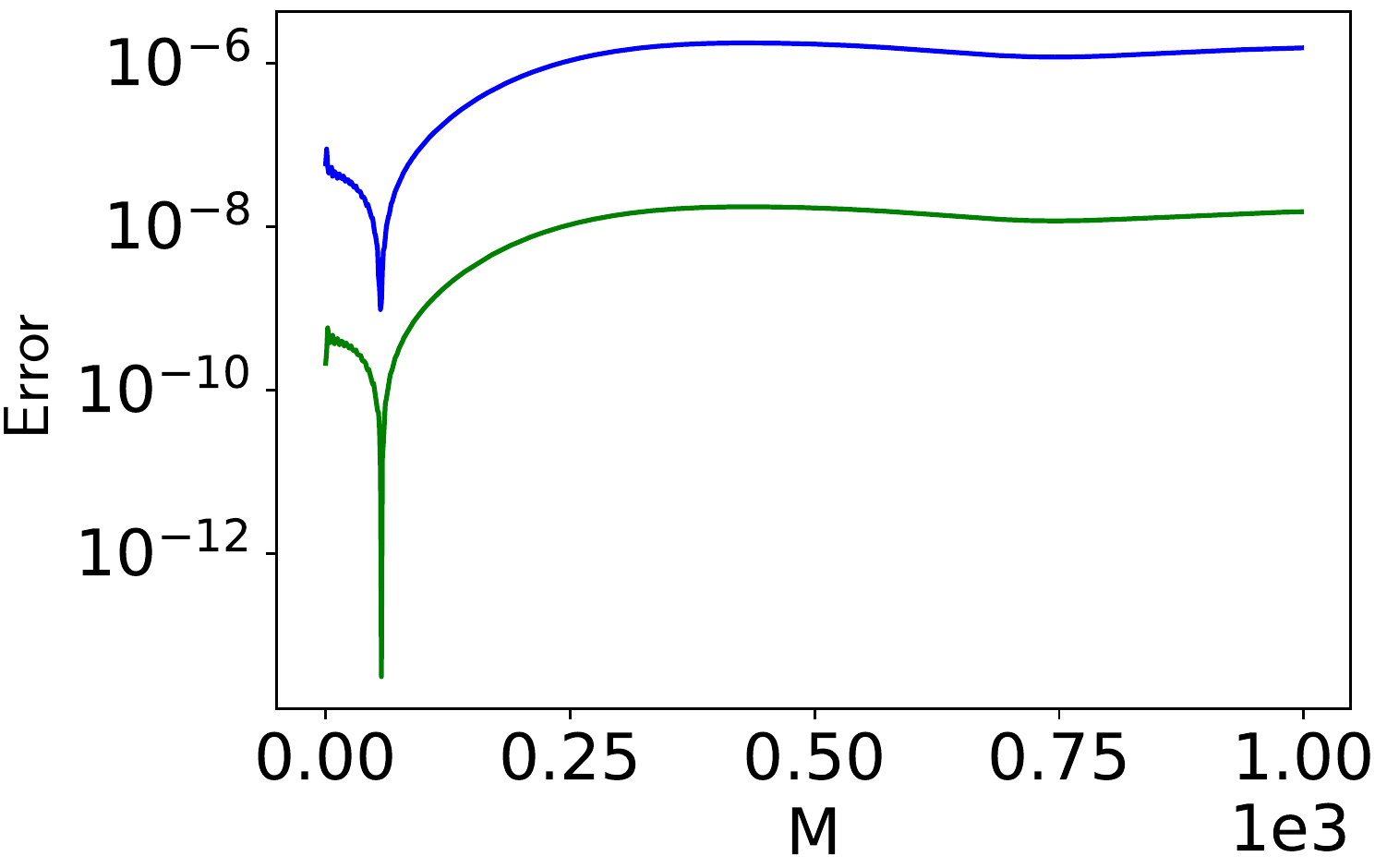} 
\includegraphics[width=3in]{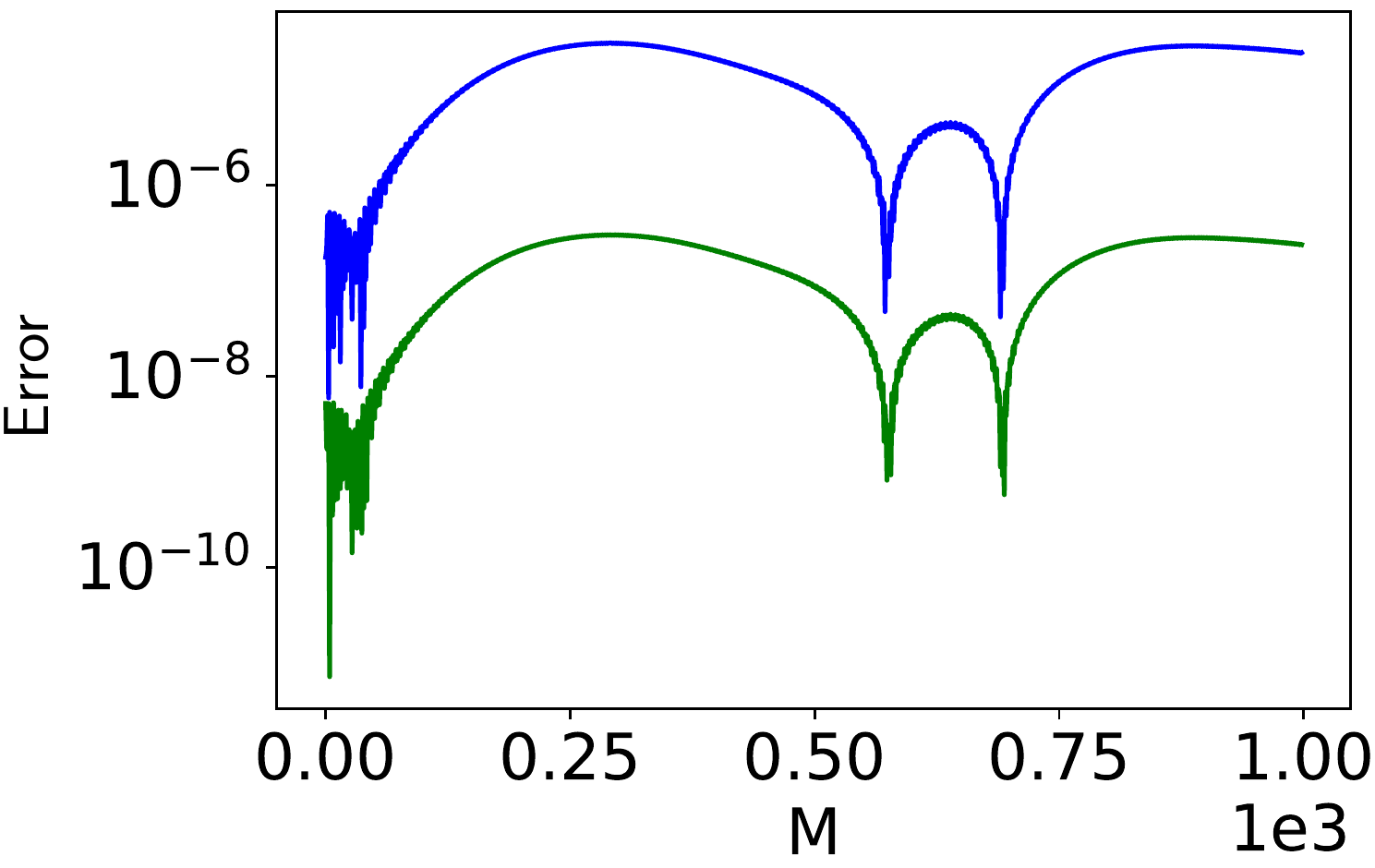}\\
\includegraphics[width=3in]{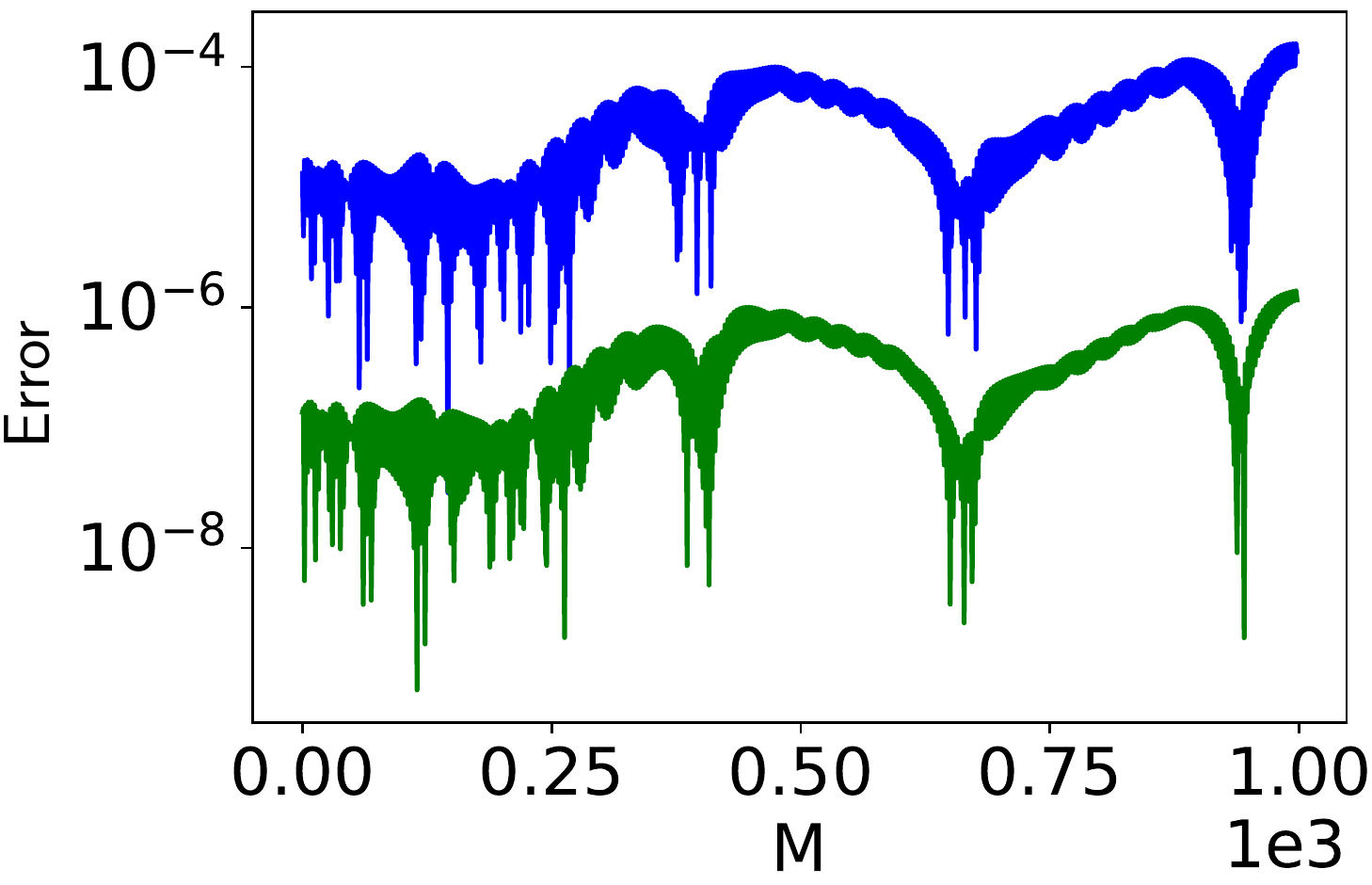}
\includegraphics[width=3in]{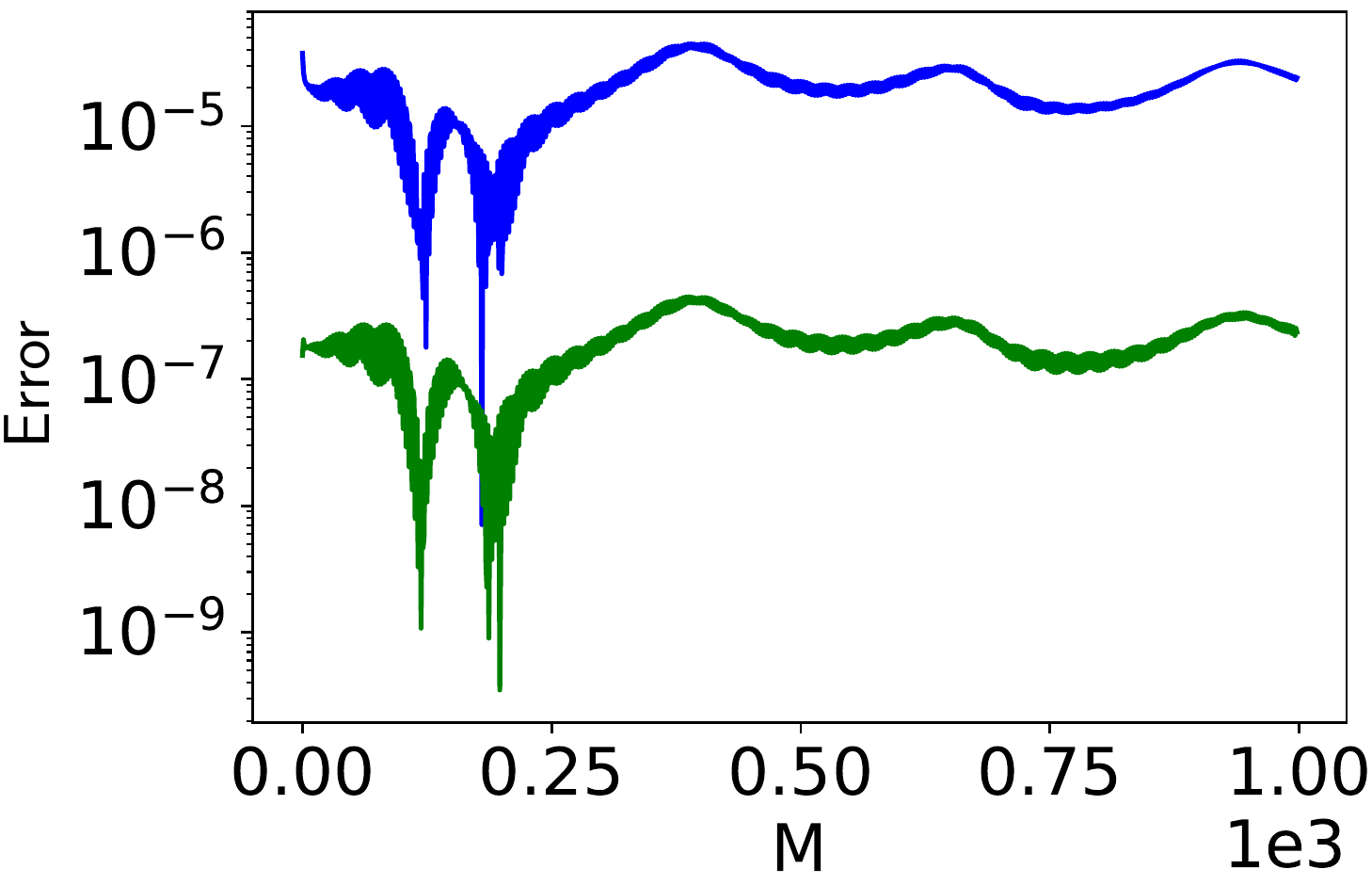}
\caption{Positional errors for the first $1000M$ of the orbit in Figure \ref{fig:Kerr_offaxis} in the $t$ (top left), $r$ (top right), $\theta$ (bottom left), and $\phi$ (bottom right) coordinates for a time step of $\delta=1$ (blue) and $\delta=0.1$ (green). The errors are computed as the fractional difference between the integrated value against a reference computation with $\delta=0.01$. The second order accurate method was chosen for these calculations, and thus the positional errors between the $\delta=1$ and $\delta=0.1$ curves are separated by a factor of $\sim 10^2$ as expected.}
\label{fig:pos}
\end{figure*}

We further test FANTASY for orbits that are neither circular or confined to the equatorial plane and confirmed that all four constants of the motion are conserved over long integration times (Figure \ref{fig:Kerr_offaxis}). In Figure \ref{fig:convergence}, we show the error in the Carter's constant as a function of stepsize, $\Delta t$, and demonstrate that the prebuilt second order and fourth order solvers obey the required rate of convergences. While symplectic integrators guarantee that the error in the conserved quantities are bounded, positional errors are not similarly privileged. In Figure \ref{fig:pos}, we show the positional errors of the integrations shown in Figure \ref{fig:Kerr_offaxis}. The positional errors are computed as
\beq
\textrm{Error} = \frac{x_n - x_r}{x_r} \; ,
\eeq
where $x_n$ is the integrated position with $\delta=1$ and $\delta=0.1$ while $x_r$ is a reference value that is computed using a high resolution timestep of $\delta=0.01$.

To test FANTASY for orbits with a more complicated metric possessing no non-zero components, we repeat our integration with in the Cartesian Kerr-Schild (KS) coordinates. The coordinate transformation from the BL coordinates to the KS coordinates are described in Appendix A. The resulting orbits computed in the two coordinate systems match each other. We transformed the results of our previous integration in BL coordinates to KS coordinates and overlaid it on top of the KS orbit in Figure \ref{fig:Kerr_offaxis}.

\section{Geodesics of the Kerr-Sen Metric}  \label{sec:Kerr-Sen}
\begin{figure*}
\centering
\includegraphics[width=1.7in]{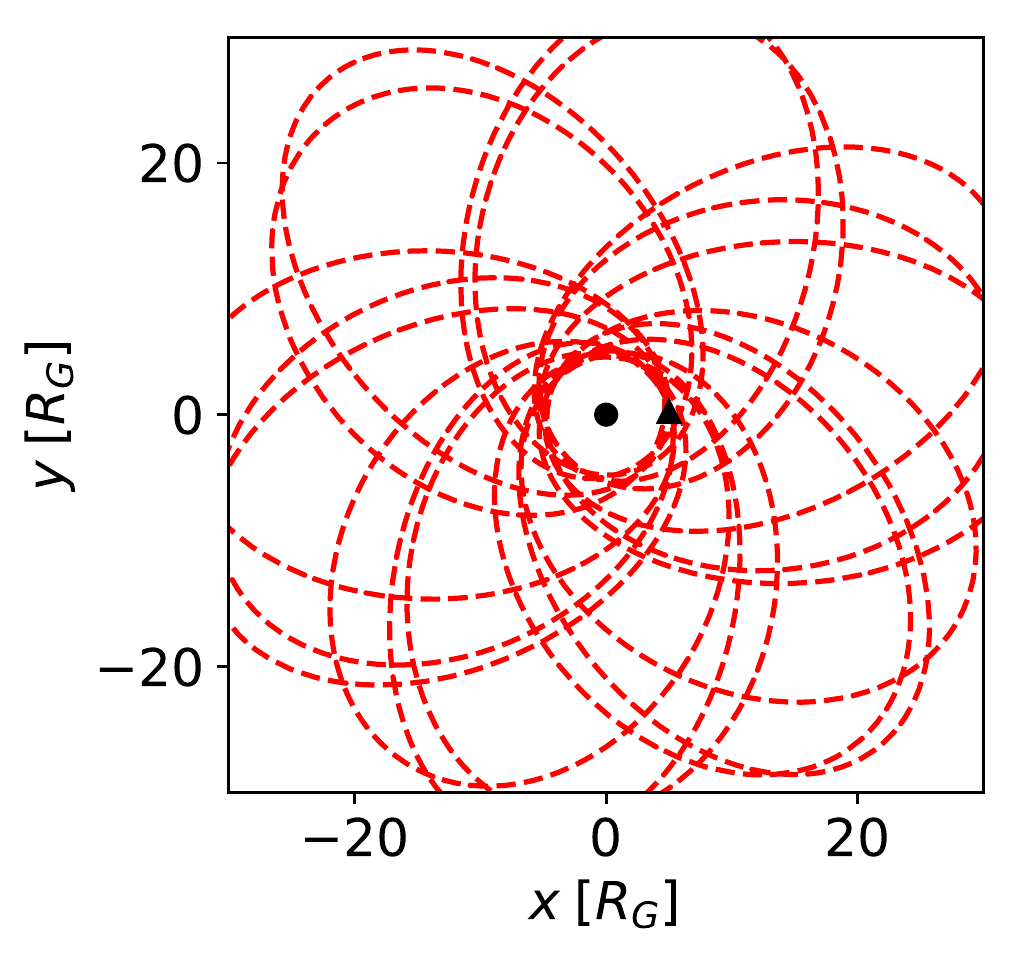} 
\includegraphics[width=1.7in]{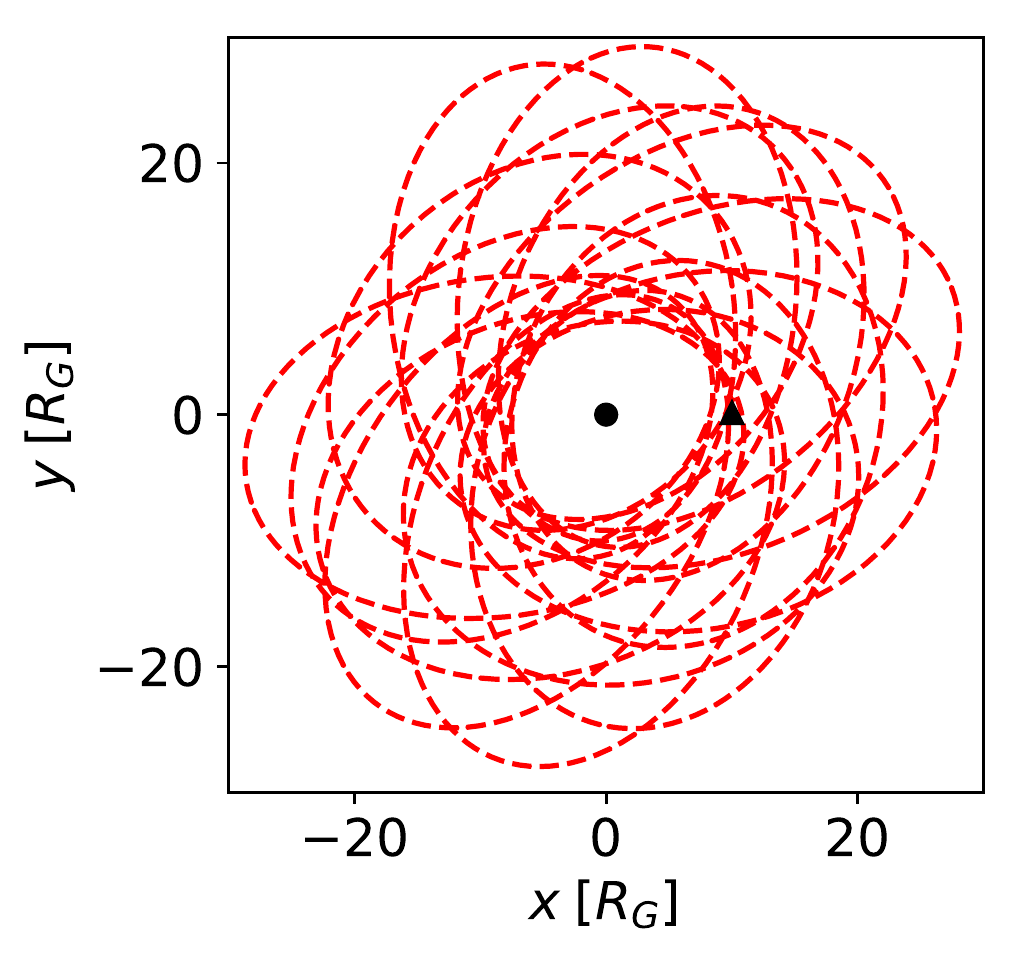} 
\includegraphics[width=1.7in]{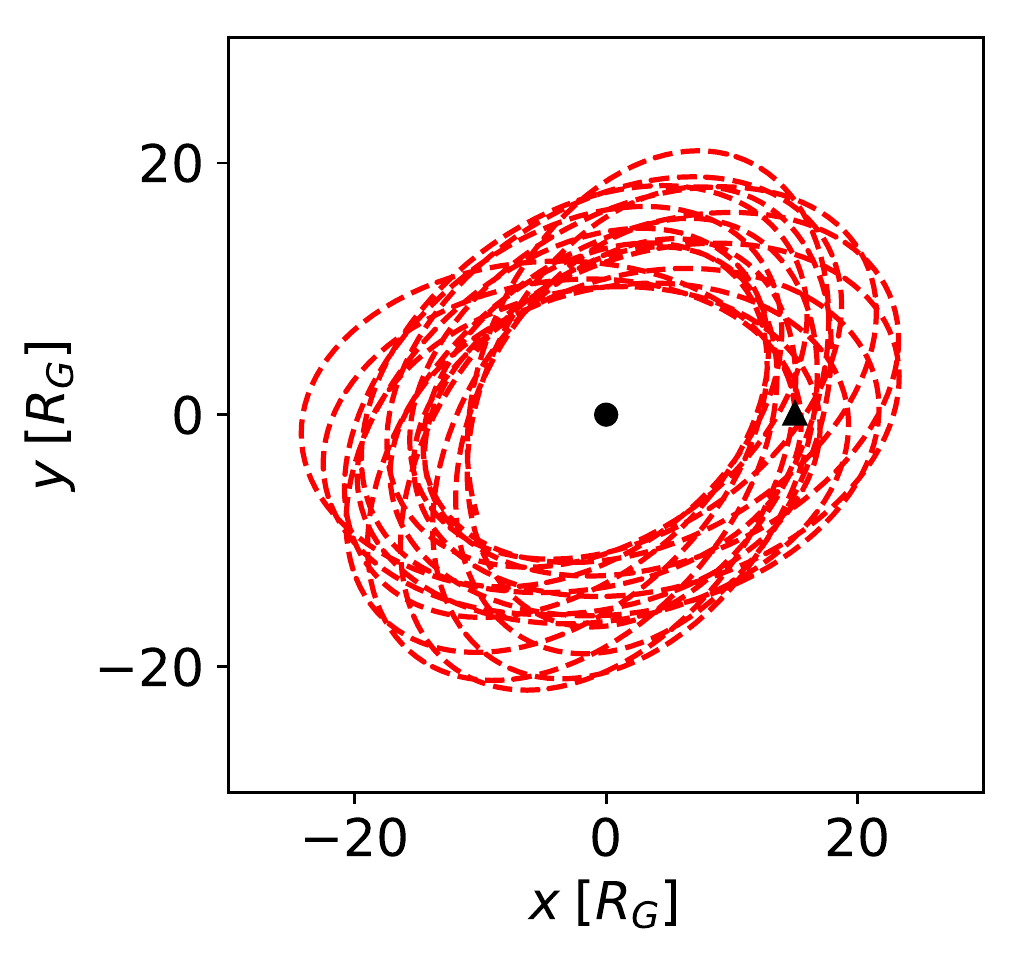} 
\includegraphics[width=1.7in]{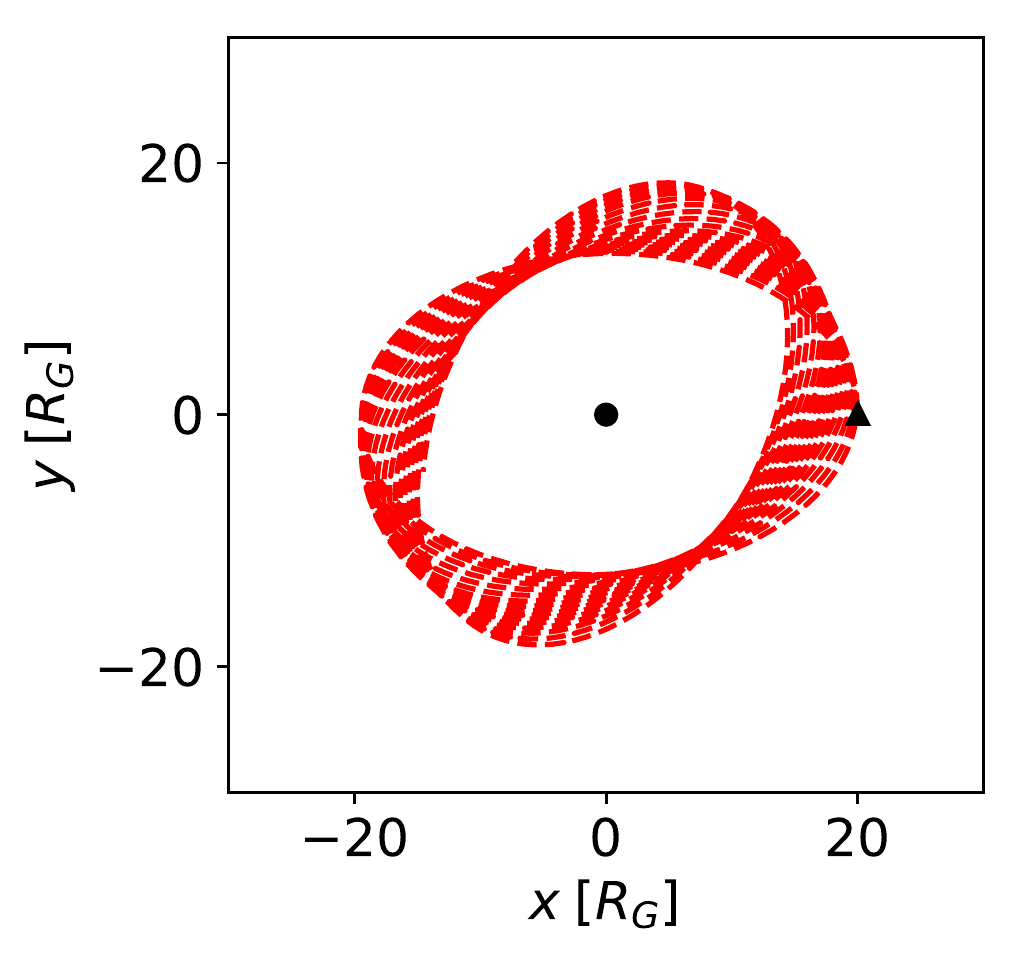} \\
\includegraphics[width=1.7in]{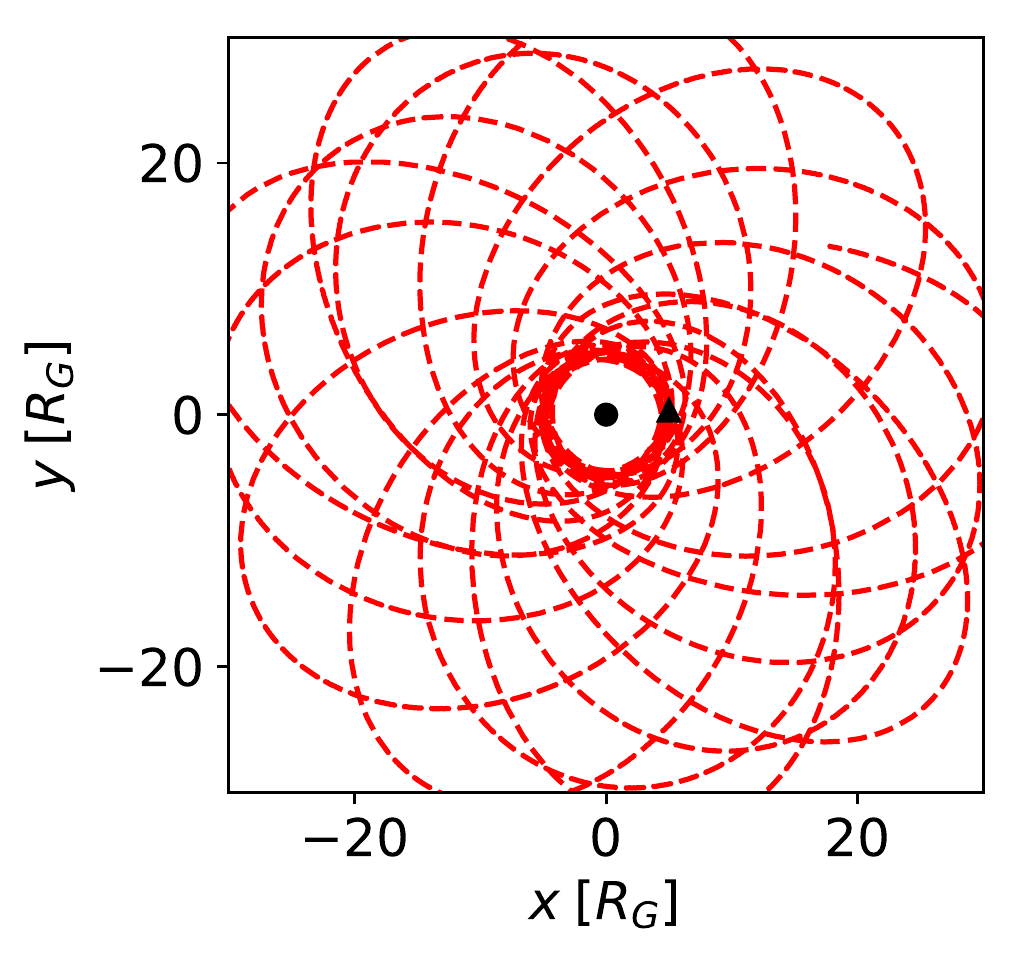} 
\includegraphics[width=1.7in]{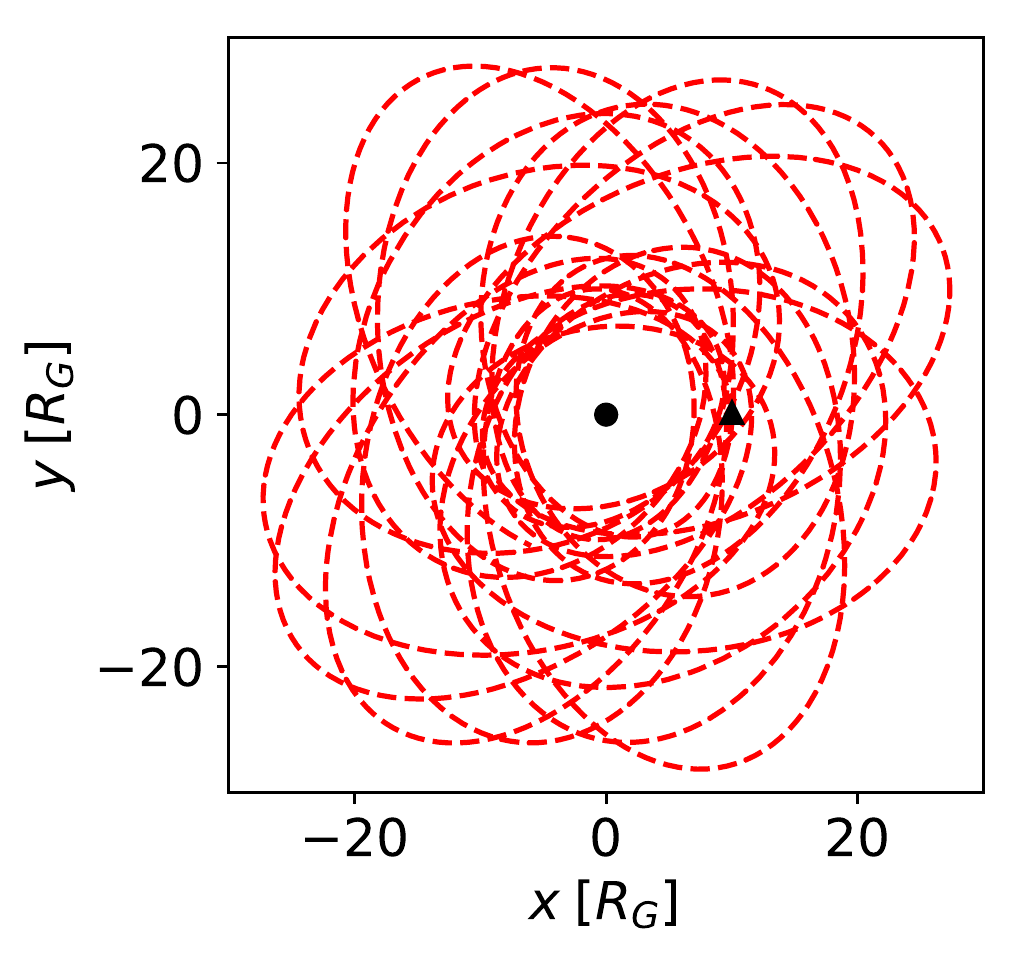} 
\includegraphics[width=1.7in]{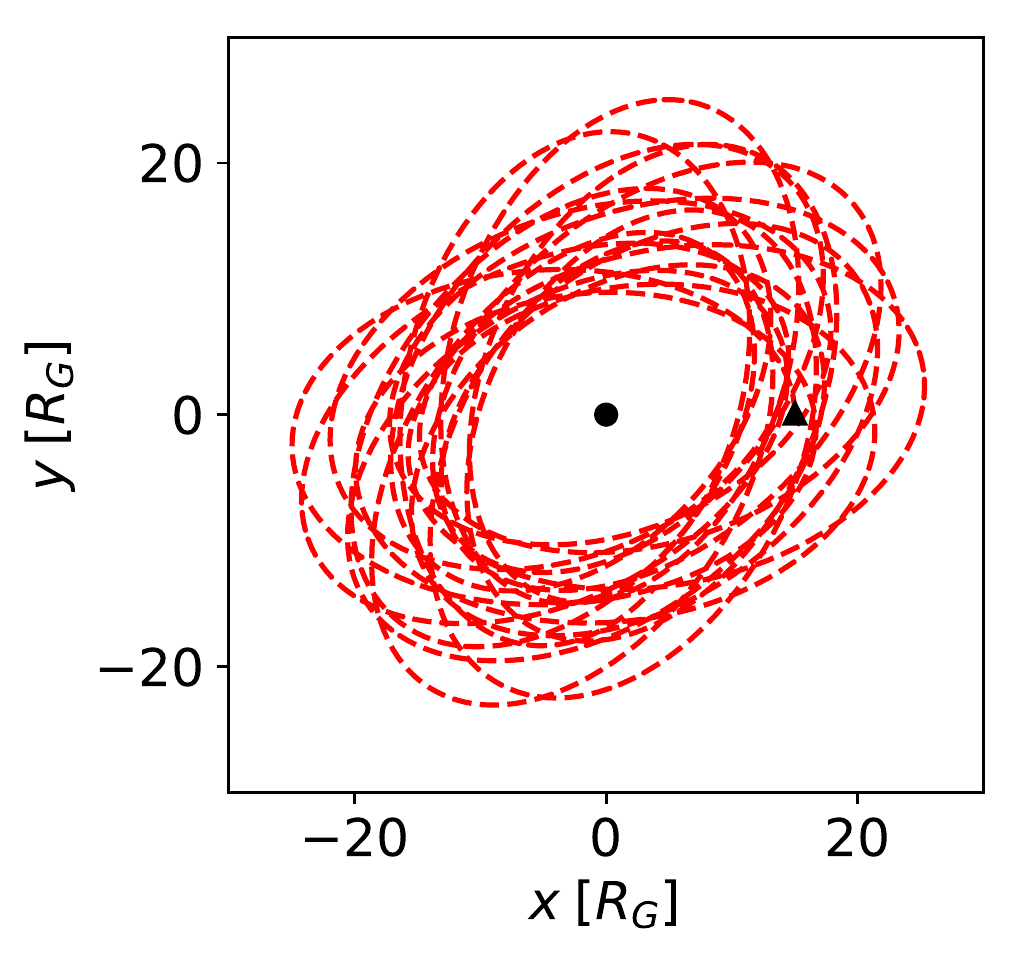} 
\includegraphics[width=1.7in]{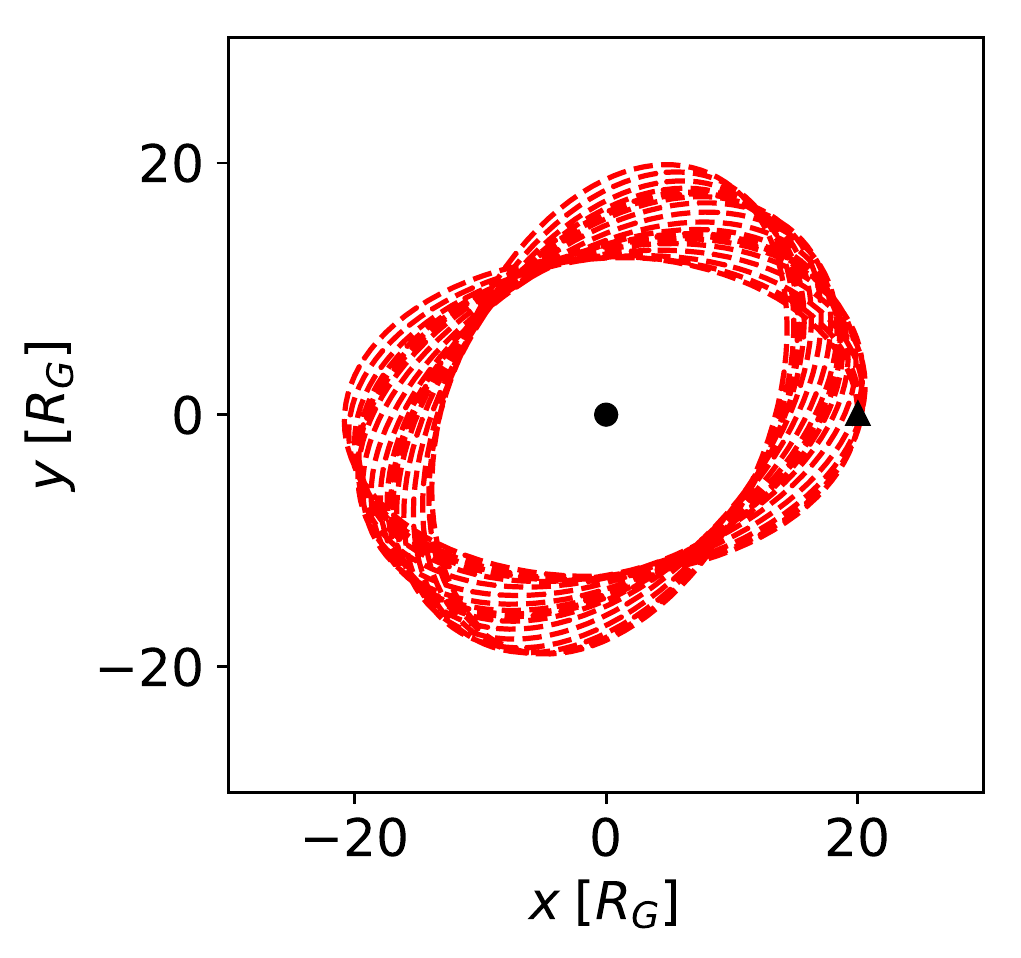} 
\caption{Geodesics around a Kerr-Sen black hole with $a=0.5$ and $b=0.9$ (top) and the geodesics starting from the same point with the same energy and angular momentum around a Kerr black hole with the same spin (bottom) projected to the equatorial plane. From the leftmost to the rightmost panels, we show geodesics with initial $r=\{5M,10M,15M,20M\}$. At small $r$, the Kerr-Sen geodesics deviate wildly from those around a Kerr black hole, but they quickly converge for geodesics orbiting further away.}
\label{fig:Kerr-Sen}
\end{figure*}

As an example application for FANTASY, we study geodesics of the Kerr-Sen metric. The Kerr-Sen metric is a black hole solution of an effective field theory corresponding to the low energy limit of heterotic string theory. In Boyer-Lindquist-like coordinates $\{ t, r, \theta, \phi \}$, the Kerr-Sen metric components in the Einstein frame read \citep{Sen},
\begin{align}
g_{tt} &= -\left( 1 - \frac{2 M r}{\Sigma} \right) \; , \\
g_{rr} &= \frac{\Sigma}{K} \; , \\
g_{\theta \theta} &= \Sigma \; , \\
g_{\phi \phi} &= \left( \Sigma + a^2 \sin^2\theta + \frac{2 M r a^2 \sin^2\theta}{\Sigma} \right) \sin^2 \theta  \; , \\
g_{\phi t} &= - \frac{2 M r a}{\Sigma} \sin^2 \theta \; ,
\end{align}
where $M$ is the mass of the black hole, $a$ the spin parameter, 
\begin{align}
\Sigma &= r(r+2b) + a^2 \cos^2 \theta \; , \\
K &=r(r+2b) - 2Mr +a^2 \; ,
\end{align}
and $b$ a scalar field. The Kerr-Sen metric reduces to the Kerr metric when $b=0$. Further, while the scalar field of the Kerr-Sen metric produces an electric charge of
\beq
Q = \sqrt{2 M b} \; ,
\eeq
we will only study the motion of electrically neutral particles. In Figure \ref{fig:Kerr-Sen}, we provide the integration of some sample geodesics in the Kerr-Sen metric with $a=0.5$ and $b=0.9$, and compare them with the analogous geodesics around a Kerr black hole that starts at the same point and possess the same energy and angular momentum. While the deviations from Kerr is large at small radii, the Kerr and Kerr-Sen geodesics converge rapidly for particles even slightly far away from the black hole. 

\section{Conclusions} \label{sec:conclude}
We present FANTASY, a new tool for the integration of geodesics in arbitrarily curved manifolds. FANTASY is designed to be user-friendly and works 'out of the box', and owing to our integration scheme and implementation of automatic differentiation, only requires the user to input the metric and initial conditions for the geodesics. The integration scheme we employ is symplectic, and thus possess bounded errors in the conserved quantities. FANTASY also comes with an automatic Jacobian calculator that allows for coordinate transformations to be computed automatically via our automatic differentiation module.

FANTASY is well suited to solve astrophysical problems like the motion of photons around non-Kerr black holes or other exotic objects with complicated metrics whose derivatives are not easy to compute, and thus can be used, for example, to aid the modeling efforts of experiments such as the EHT or employed in pulsar timing computations for exotic pulsar binaries. FANTASY can also be modified to incorporate numerical spacetimes. The most straightforward method to achieve this is by combining it with an algorithm that gives an analytical representation of a numerical metric (through, e.g., spline or polynomial fitting).

FANTASY is open source, and is available at https://github.com/pierrechristian/FANTASY\citep{code}. 

\section{Acknowledgements}
We acknowledge Ian Weaver for discussions on the use of dual numbers for automatic differentiation and Gabriele Bozzola for discussions on code optimization.

\appendix

\section{Coordinate Transforms and Spherical Kerr-Schild Coordinates}
\label{sec:transforms}

In this appendix, we provide the coordinate transformation between
the Cartesian Kerr-Schild (KS) coordinates, $(t_{\rm{KS}},x,y,z)$, and Boyer-Lindquist (BL) coordinates, $(t,r,\theta,\phi)$. The steppingstone for the transformation is the spherical KS
coordinate system, $(t_{\rm{KS}},r,\bar{\theta},\bar{\phi})$, which has a simple relations to both the Cartesian KS and the BL coordinates. The radial component of the spherical KS coordinate is simply the $r$ of the BL coordinates, which can be written explicitly in terms of Cartesian KS coordinates as,
\begin{align}
  r &= \left[\frac{R^2 - a^2 +
                   \sqrt{(R^2 - a^2)^2 + 4 a^2 z^2}}{2}\right]^{1/2} \; ,
  \label{eq:r_ex}
\end{align}
with $R \equiv \sqrt{x^2 + y^2 + z^2}$.
The KS polar angle $\bar{\theta}$ and KS azimuthal angle $\bar{\phi}$ are defined by
\begin{align}
  \cos\bar{\theta} &= z / r \; , \\
  \tan\bar{\phi}   &= (xa + yr)/(xr - ya) \; .
\end{align}
It is easy to verify that the inverse transformations from spherical
to Cartesian KS are
\begin{align}
  x &= (r\cos\bar{\phi} + a\sin\bar{\phi})\sin\bar{\theta} \; , \\
  y &= (r\sin\bar{\phi} - a\cos\bar{\phi})\sin\bar{\theta} \; , \\
  z &= r\cos\bar{\theta} \; .
\end{align}
Note that when $a\ne0$, intuition from flat three-dimensional space is no longer valid. For example, a particle starting at the equator with Spherical KS coordinates $(0,r_0,0,0)$ possesses the initial condition $(0,r_0,-a,0)$ in Cartesian KS instead of the $(0,r_0,0,0)$ that one would expect from the usual spherical to Cartesian coordinates transformation in flat spacetime.

As with the radial coordinate, the polar angle of the spherical KS is also identical to the polar angle of the BL coordinate system,
\beq
\theta = \bar{\theta} \; .
\eeq
The time and azimuthal coordinates are related by the following,
\begin{align}
  t_{\rm{KS}}    &= t + \int\frac{2Mr}{\varDelta}\mathrm{d}r, \\
  \bar{\phi} &= \phi + \int\frac{a}{\varDelta}\mathrm{d}r.
\end{align}
The above integrals have analytical solutions,
\begin{align}
  t_{\rm{KS}}  - t &=  \frac{2M^2}{\sqrt{a^2-M^2}}\tan^{\!-1}\!\left(\frac{r - M}{\sqrt{a^2-M^2}}\right) + M\ln(\varDelta/M^2) + \mathrm{constant} \; , \\
  \bar{\phi} - \phi &=  \frac{a}{\sqrt{a^2-M^2}}  \tan^{\!-1}\!\left(\frac{r-M}{\sqrt{a^2-M^2}}\right) + \mathrm{constant} \; .
\end{align}
Since we expect $a < M$ for astrophysical black holes,
$\sqrt{a^2-M^2}$ is purely imaginary.
Let $h \equiv \sqrt{M^2 - a^2}$ be real and $r_\pm \equiv M \pm h$ be
the outer and inner event horizons, respectively, we can then use the
properties of complex arctangent and complex natural log to rewrite,
\begin{align}
  \frac{1}{ih}\tan^{\!-1}\!\left(\frac{r - M}{ih}\right) &=
  \frac{1}{2h}\left[\ln\left(\frac{r - r_+}{r - r_-}\right) +
    i\pi(2n + 1)\right] \; ,
\end{align}
for an arbitrary $n$.
Although the above equation is complex, we can choose the integration
constants in ways that the final coordinate transformations are real. This integration constant can be obtained by setting $t_{\rm{KS}}  = t$ and $\bar{\phi} =
\phi$ at the initial radius $r_0$ so that the transformations become,
\begin{align}
  t_{\rm{KS}}  - t &= M\left[
    \frac{r_+}{h} \ln\left(\frac{r - r_+}{r_0 - r_+}\right)
  - \frac{r_-}{h} \ln\left(\frac{r - r_-}{r_0 - r_-}\right)
  \right] \; , \\
  \bar{\phi} - \phi &= \frac{a}{2h}\left[
    \ln\left(\frac{r - r_+}{r_0 - r_+}\right)
  - \ln\left(\frac{r - r_-}{r_0 - r_-}\right)
  \right] \; .
\end{align}
The right hand sides of the above equations are singular at
the horizon. This comes no surprise because there are geodesics that can pass the
horizon at finite $t_{\rm{KS}} $ and $\bar{\phi}$ in KS but will be mapped to infinity
in the $t$ and $\phi$ of BL coordinates.

Transformations of tensorial objects from the Cartesian KS to the BL coordinates and vice-versa can then be obtained as per usual by utilizing the Jacobian, $\mathbf{J}$. For example, for the vector $\mathbf{v}$ and dual vector $\mathbf{w}$,
\begin{align}
\bar{\mathbf{v}} &= \mathbf{J} \mathbf{v} \; , \\
\bar{\mathbf{w}} &= \mathbf{J}^{-1} \mathbf{w} \; , 
\end{align}
where overbar denotes quantities in the Cartesian KS coordinates and $\mathbf{J}^{-1}$ is the inverse Jacobian obtained by inverting the Jacobian matrix. 

\bibliography{fantasybib.bib}

\end{document}